\documentclass{article}

\PassOptionsToPackage{numbers}{natbib}
\usepackage[preprint]{neurips_2025}

\usepackage[utf8]{inputenc} %
\usepackage[T1]{fontenc}    %
\usepackage{hyperref}       %
\usepackage{tcolorbox}
\usepackage{url}            %
\usepackage{booktabs}       %
\usepackage{amsfonts}       %
\usepackage{nicefrac}       %
\usepackage{microtype}      %
\usepackage{xcolor}         %
\usepackage{graphicx}
\usepackage{cleveref}
\usepackage{subcaption}
\usepackage{multirow}
\usepackage{xspace}
\usepackage{wrapfig}
\newcommand{\red}[1]{{\color{red}#1}}
\newcommand{\green}[1]{{\color{green}#1}}

\newcommand{\anja}[1]{\textbf{\color{blue}[AR: #1]}}
\newcommand{\marcel}[1]{\textbf{\color{green}[MK: #1]}}

\newcommand{\ourname}{SIMBA}
\newcommand{\FAVC}{FakeAVCeleb}
\newcommand{\DS}{DeepSpeak~v1}
\newcommand{\DF}{DeepFake}
\newcommand{\FS}{FaceSwap}
\newcommand{\FSGAN}{FS-GAN}
\newcommand{\wl}{Wav2Lip}

\newcommand{\wlS}{Wav2Lip Split}

\newcommand{\etal}{\textit{et al}\onedot}

\title{{\vspace{-3mm}\includegraphics[scale=0.1]{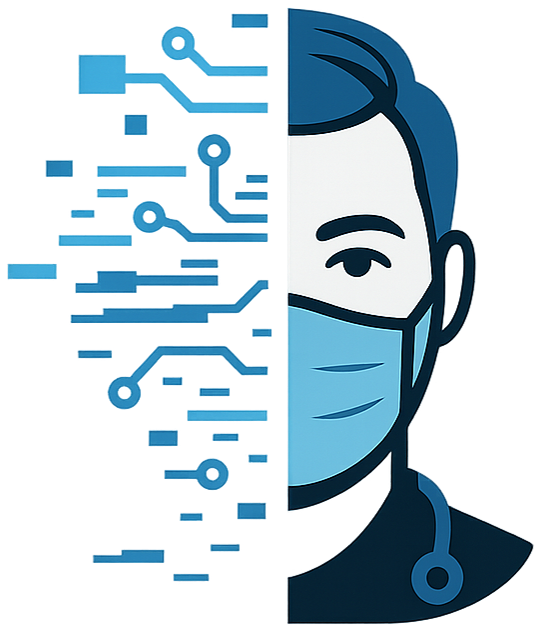}} DeepFake Doctor: \\ Diagnosing and Treating Audio-Video Fake Detection}

\author{
  Marcel Klemt \thanks{Equal contribution} \quad
  Carlotta Segna\footnotemark[1] \quad
  Anna Rohrbach \\
  TU Darmstadt \& Hessian.AI \\
  \texttt{\{marcel.klemt, carlotta.segna, anna.rohrbach\}@tu-darmstadt.de}
}

\begin{document}

\maketitle

\begin{abstract}

Generative AI advances rapidly, allowing the creation of very realistic manipulated video and audio. This progress presents a significant security and ethical threat, as malicious users can exploit \DF\ techniques to spread misinformation. Recent \DF\ detection approaches explore the multimodal (audio-video) threat scenario. In particular, there is a lack of reproducibility and critical issues with existing datasets - such as the recently uncovered silence shortcut in the widely used \FAVC\ dataset. Considering the importance of this topic, we aim to gain a deeper understanding of the key issues affecting benchmarking in audio-video \DF\ detection. We examine these challenges through the lens of the three core benchmarking pillars: datasets, detection methods, and evaluation protocols.
To address these issues, we spotlight the recent \DS\ dataset and are the first to propose an evaluation protocol and benchmark it using SOTA models. We introduce SImple Multimodal BAseline (SIMBA), a competitive yet minimalistic approach that enables the exploration of diverse design choices. We also deepen insights into the issue of audio shortcuts and present a promising mitigation strategy. Finally, we analyze and enhance the evaluation scheme on the widely used \FAVC\ dataset. Our findings offer a way forward in the complex area of audio-video DeepFake detection.
\end{abstract}

\section{Introduction}
\label{sec:intro}

Over the past decade, Generative AI has advanced rapidly, enabling the creation of highly realistic synthetic content. User-friendly applications allow anyone to generate convincing \DF s of friends or public figures using just a few seconds of footage. In 2023 alone, over 500,000 fake videos were shared globally~\cite{openfoxDeepfakesTheir}, reflecting the growing accessibility and impact of this technology. As \DF s become more prevalent, so do the risks they pose, namely misinformation and identity manipulation, political interference, and fraud. \DF s are now a widespread and serious concern, as they are often used with harmful intent. With continued progress in Generative AI across modalities, including image, video, and audio, there is a rise in multimodal (audio-visual) \DF s. %

To counter these threats, research on \DF\ detection has grown in parallel with generation techniques. Most of the prior work has explored the unimodal scenarios (predominantly video-only~\cite{zheng2021exploring, LipForensics, RealForensics}, some audio-only~\cite{audio_detection_2}). Recent literature is increasingly focused on the multimodal approaches~\cite{Bohacek_2024_CVPR, AudioVisualDissonance, mittal2020emotions, AVFF, avad} applicable to cases where only one or both modalities are manipulated. As new and increasingly complex approaches are being proposed, dataset issues are being brought to light~\cite{CircumventingShortcuts}, thus, we may ask: are we in fact making progress in audio-video \DF\ detection?

Considering the importance of this topic, \emph{we aim to get a deeper understanding of the key issues prevalent to benchmarking in audio-video \DF\ detection}. We characterize such issues with respect to the three benchmarking pillars: \emph{datasets}, (detection) \emph{methods}, and \emph{evaluation protocols}. %

\textbf{Datasets.} In order to train and evaluate multimodal \DF\ detection models, several datasets have been proposed, with the goal to capture both video and audio modality~\cite{dfdc, kodf, khalid2021fakeavceleb, deepspeak_dataset, polyglot_fake, lavdf, AVDeepFake1M, TMC_dataset, AVoiD-DF, MMDFD_dataset, AVLips}. Yet, most of these datasets suffer from one or more issues. 
(a) Some of them are not publicly available~\cite{TMC_dataset, AVoiD-DF, MMDFD_dataset} or have partial availability~\cite{polyglot_fake}. %
(b) Some datasets only offer manipulations for the visual modality, e.g., KoDF~\cite{kodf} and AVLips~\cite{AVLips}, preventing the study of diverse combinations of manipulated modalities. %
(c) Datasets like DFDC~\cite{dfdc} and AVLips~\cite{AVLips} only provide binary labels (real vs. fake) and no manipulation labels, thus not supporting the evaluation of cross-manipulation generalization, which is important for practical applicability. %
(d) Boldisor~\emph{et al.}~\cite{CircumventingShortcuts} recently uncovered the presence of \emph{shortcuts} in \FAVC~\cite{khalid2021fakeavceleb} and AV-DeepFake1M~\cite{AVDeepFake1M} manifested as leading silence which can be exploited by models without having to actually learn the task.%
(e) Popular datasets, such as \FAVC, seem to be fairly saturated~\cite{AVFF, LipForensics,RealForensics,spatiotemporalinconsistency}. Not only is it solved almost to perfection, but it also presents a shortcut.
Yet, these datasets form the foundation of this research area. To allow further successful development of multimodal \DF\ detection models and enable measuring the progress, \emph{new open high-quality benchmarks} are required. Besides, we need to deepen our understanding of the \emph{shortcut issue} and find possible mitigation strategies.

\textbf{Methods.} The benchmarking issues are further complicated by often \emph{unavailable \DF\ detection models' implementations}, as many authors do not make their models publicly available~\cite{AVFF, AVoiD-DF, Multimodaltrace}. This hinders reproducibility and often makes it impossible to compare to or build upon prior work. %

\textbf{Evaluation Protocols.}
As already mentioned above, generalization is a crucial aspect when evaluating \DF\ detection methods as they tend to overfit to the training data artifacts and do not generalize beyond~\cite{kamat2023revisiting}. Most prior work includes some form of cross-manipulation (within a dataset) or cross-dataset evaluation. But also here some issues can be found, stemming from mismatched experimental setups used across different works.

Our efforts to address the above issues are as follows. 
First, we spotlight the dataset previously unexplored for multimodal \DF\ detection, \DS~\cite{deepspeak_dataset}, and make a case for its use as a new benchmark. \DS\ contains more recent manipulation techniques, extreme head poses, and occlusions. 
We also diagnose \DS\ regarding the shortcut phenomenon.
Additionally, we revisit the popular \FAVC~\cite{khalid2021fakeavceleb} dataset, where we extend the shortcut analysis offered in~\cite{CircumventingShortcuts} to all the manipulations. We provide a manipulation-specific breakdown, showing how all the \emph{fake-audio} manipulations are impacted!  %

For our benchmarking and analysis, we introduce a new SImple Multimodal BAseline (\ourname); it features a minimalistic design with two modality encoders (audio\&video) followed by a late fusion and a classification head (supported by two unimodal helper heads). Using \ourname, we study different design choices, such as temporal sampling and augmentation strategies that can be performed at training time. We show that these have a large impact on the model's susceptibility to the shortcuts, offering a \emph{simple and promising mitigation strategy}.

Last, we revise the existing cross-manipulation evaluation protocol on \FAVC. First, we uncover blind spots (some manipulations were left out!), and manipulation-leakage in the established leave-one-out protocol. We propose new protocols (method and family splits) with different levels of generalization challenges. We are, to the best of our knowledge, the first to present a similar evaluation protocol for \DS\ and benchmark SOTA models against it. Lastly, we study cross-dataset generalization with our two benchmarks, spanning the full spectrum of manipulations.

\section{Related Work}
\label{sec:related_work}

\textbf{Multimodal \DF\ Detection Datasets.} 
The DeepFake detection community has introduced various datasets for training and evaluation. These typically feature two video manipulation ``families'': \emph{lip synthesis} (lip area only) and \emph{face animation} (entire face). For audio, the most common manipulation is \emph{Text-To-Speech (TTS)}.
Datasets are generally categorized into \emph{video-only} and \emph{audio-video} (multimodal) manipulations.
Video-only datasets include FaceForensics++~\cite{FaceForensics}, ForgeryNet~\cite{ForgeryNet}, DF-Platter~\cite{DFPlatter}. Besides AVLips~\cite{AVLips}, and KoDF~\cite{kodf}, contain various video manipulations but only real audio; nonetheless, they are still used in the domain of multimodal \DF\ detection. 
Audio-video datasets include \FAVC~\cite{khalid2021fakeavceleb}, \DF\ Detection Challenge~\cite{dfdc} (DFDC), \DS~\cite{deepspeak_dataset}, Lav-DF~\cite{lavdf}, PolyGlotFake~\cite{polyglot_fake}, and AV-DeepFake1M~\cite{AVDeepFake1M}. 
DFDC uses various manipulations but lacks manipulation-specific labels, thus not supporting cross-manipulation evaluation. Lav-DF and AV-DeepFake1M focus on DeepFake localization (detecting manipulated regions); each uses just one lip synthesis manipulation. \FAVC\ and \DS\ offer annotated manipulation types, enabling cross-manipulation evaluation. Further, they both include lip synthesis and face animation as generation techniques. 
PolyGlot introduces multilingual \DF\ scenarios but suffers from missing and low-quality samples.

For our work, we focus on the \FAVC\ and \DS\ datasets, as they present different labeled types of manipulation in both audio and video modality.

\textbf{\DF\ Detection Methods.} The \DF\ (DF) detection landscape includes both \textbf{unimodal} and \textbf{multimodal} approaches. Among {Unimodal detection} approaches, video methods like LipForensics~\cite{LipForensics} and RealForensics~\cite{RealForensics} detect artifacts in lips or frames, achieving strong results on FaceForensics++. Audio-based methods transform audio into log spectrograms~\cite{audio_detection_1, audio_detection_3} and use classifiers or capsule networks~\cite{audio_detection_2} to capture features and temporal dynamics. %
{Multimodal detection} integrates audio and video, typically detecting misalignment between the two modalities~\cite{VideoFaceHomogeneity, AudioVisualDissonance, AVLips, CircumventingShortcuts, avad, speechforensics, Zhou_2021_ICCV}, phoneme-viseme mismatches~\cite{Phoneme-Viseme-Mismatch} or identity shifts~\cite{UnsupervisedDFDectionIntraCross}. AVoiD-DF~\cite{AVoiD-DF} uses a Temporal-Spatial Encoder and cross-modal classifier to identify inconsistencies. AVFF~\cite{AVFF} reconstructs masked inputs from complementary modalities. Multimodaltrace~\cite{Multimodaltrace} fuses features via mixer layers and predicts modality labels using a multilabel head.
\section{Dataset Analysis}

\begin{figure}
     \centering
     \begin{subfigure}[b]{0.48\textwidth}
        \centering
        \includegraphics[width=\linewidth]{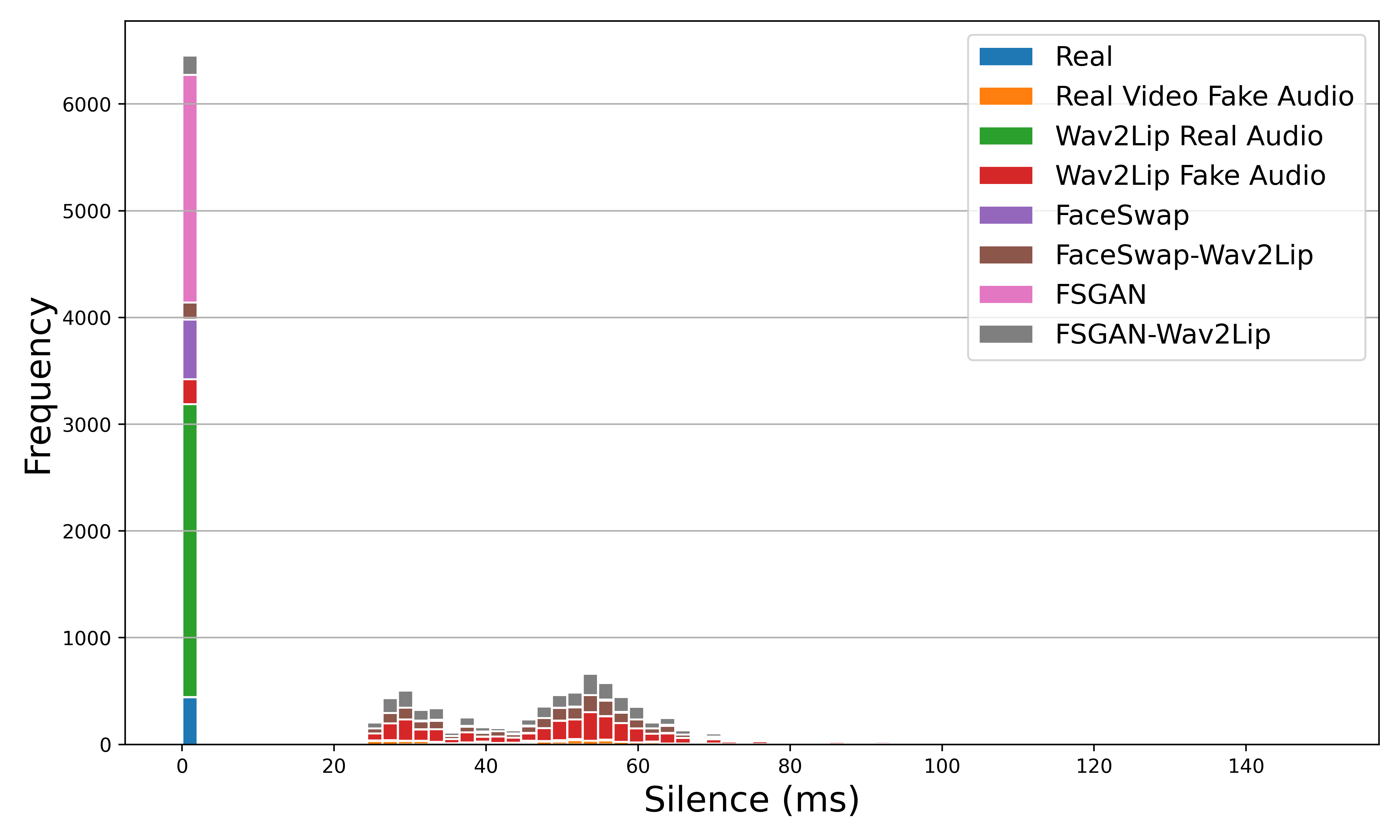}
        \caption{FAVC leading silence distribution}
        \label{fig:favc_leadingsilence}
     \end{subfigure}
     \begin{subfigure}[b]{0.48\textwidth}
        \centering
        \includegraphics[width=\linewidth]{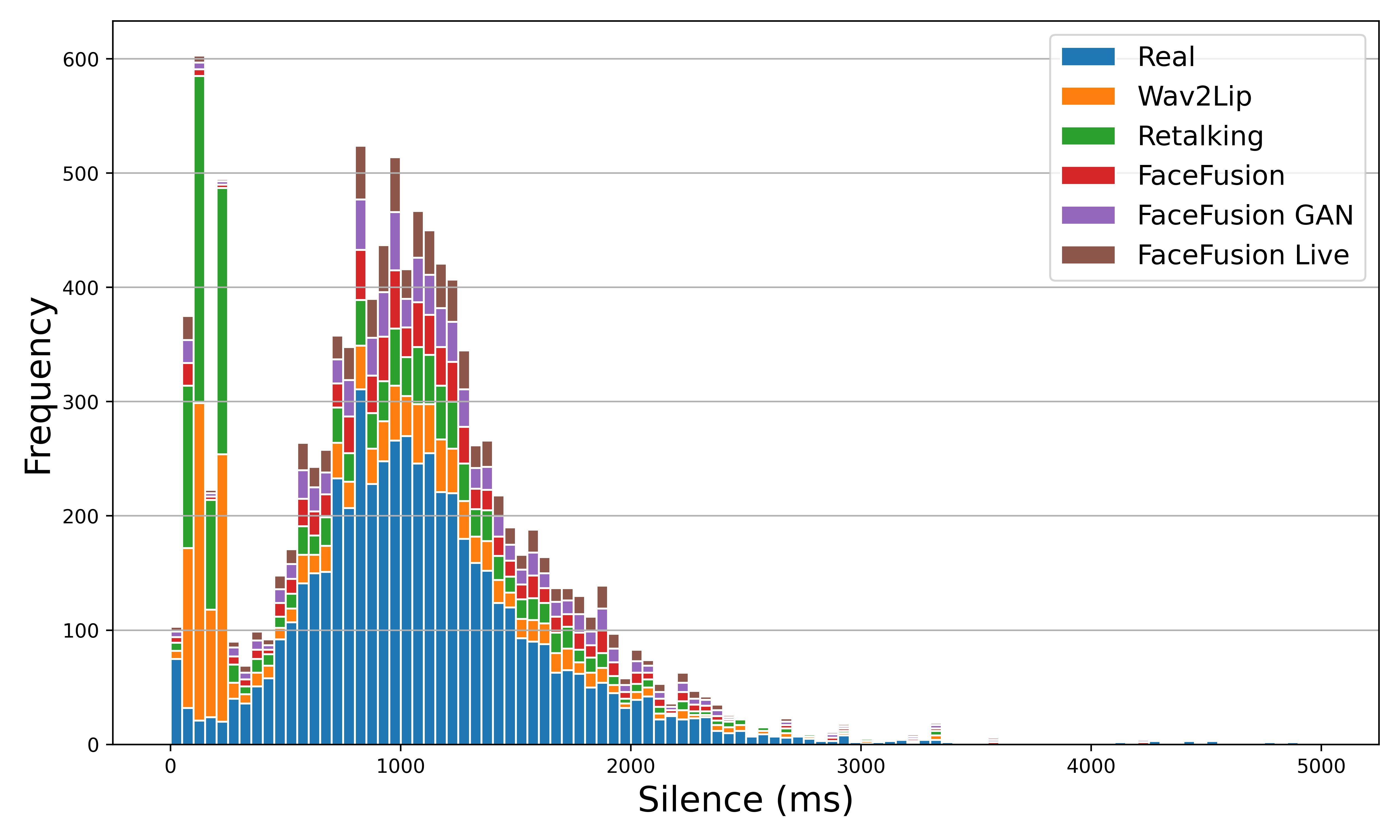}
        \caption{\DS\ leading silence}
        \label{fig:ds_leadingsilence}
     \end{subfigure}
            \caption{Leading silence distribution of the FAVC and \DS\ datasets.}
        \label{fig:favc_deepspeak_leadingsilence}
\end{figure}

\label{sec:data_analysis}

\subsection{\FAVC}
\label{sec:favc_description}

\textbf{Dataset Composition.}
The dataset \FAVC~\cite{khalid2021fakeavceleb} is composed of three different video manipulation techniques and one audio manipulation technique. For video, Wav2Lip~\cite{wav2lip2020} is used as a lip synthesis technique, whereas FaceSwap~\cite{FaceSwap} and FSGAN~\cite{nirkin2019fsgan} are employed for entire face animation. Audio fakes are synthesized using a single method, SV2TTS~\cite{SV2TTS}. Both the lip synthesis and the face animation techniques are either applied alone, generating fakeVideo-realAudio or realVideo-fakeAudio samples, or in combination (fakeVideo-fakeAudio samples). Overall, \FAVC\  contains 21k videos from 500 identities, originally selected from the VoxCeleb2 dataset~\cite{Chung18b}. We provide detailed information about the manipulation distribution in~\Cref{supp:protocol}. 

\textbf{Shortcut Issue.}
As recently uncovered, at least part of the \FAVC\ dataset suffers from a \emph{leading silence shortcut}. Boldisor~\emph{et al.}~\cite{CircumventingShortcuts} disclosed that fakeVideo-fakeAudio samples include several milliseconds of silence at the beginning not present in real samples. To further analyze this, we compute the silence by setting a threshold to $20db$ and consider it only if it lasts at least $20ms$. We extend the previous analysis with a manipulation-specific breakdown of leading silences, displayed in~\Cref{fig:favc_leadingsilence}.
It can be seen that the leading silence is present in all manipulations that involve \emph{fake audio}. As shown by~\cite{CircumventingShortcuts}, supervised models ``latch on'' this shortcut to differentiate between real and fake samples, while self-supervised approaches, which do not see fake samples during training, are agnostic to the leading silence. Since FakeAVCeleb is one of the most established datasets in this research field, this discovery poses the question of whether supervised (especially multimodal) models can still use this dataset and whether the results proposed so far still hold. We discuss our baseline approach in \Cref{sec:method}, where we take the shortcut issue into account.

\subsection{DeepSpeak v1}%
\label{sec:deepspeak_description}

\textbf{Dataset Composition.}
The \DS~\cite{deepspeak_dataset} dataset is composed of five different video and one audio generation technique. Specifically, for video manipulations, it utilizes FaceFusion~\cite{facefusion}, FaceFusion+GAN~\cite{facefusion_gan}, FaceFusion Live (FaceFusion but simulating a live streaming environment), Wav2Lip~\cite{wav2lip2020} and Video Retalking~\cite{videoretalking_ds}. ElevenLabs’ voice cloning API~\cite{elevenlabsElevenLabsFree} is used for generating audio fakes. Unlike \FAVC, the audio manipulation is only present when a lip synthesis technique is used. The face animation manipulations include solely real audio. Overall, the dataset encompasses 13k videos from 220 identities. More details can be found in \Cref{supp:protocol}. %
%
%
%
%
%
%
%
%
%
%
%
%
%
%
%
%

\textbf{Shortcut Issue.}
First, we analyze whether leading silence is also present in \DS. The silence is detected by setting the threshold to $20db$ and considering silence only if it lasted at least $20 ms$, as for \FAVC. 
~\Cref{fig:ds_leadingsilence} shows the leading silence histogram over the dataset. Here, the majority of samples have a leading silence of various lengths. Yet, the distribution is balanced between real and fake and along the individual manipulations. The only exception is the first four bins where \wl\ and Retalking dominate, raising the possibility of a shortcut. 

Both datasets are heavily skewed to the video modality, only featuring a single audio manipulation. Further, they share one manipulation, namely Wav2Lip~\cite{wav2lip2020}. 
We report our findings about the strength of the leading silence shortcut in both datasets in \Cref{sec:shortcut_results}.
Further, we include the generalization evaluation for cross-manipulation and cross-dataset in \Cref{sec:cross_manipulation_results} and \ref{sec:cross_dataset_results}, respectively.

%
%
%
%
%
%

%
%
%
%
%
%
%
%
%
%
%
%
%
%
%
%

%
%
%

%

%

%
%

%

\section{Methodology}
\label{sec:method}

As previously discussed, one of the challenges we face is the lack of publicly available multimodal detection methods. Many promising models do not release their code or, when it is released, are missing components to make it executable. Thus, we opt to develop \ourname, our SImple Multimodal BAseline. \ourname\ is a multimodal model, which comprises an audio and video encoder, followed by a fusion branch, trained for the task of \DF\ detection.

\begin{figure*}[t]
\centering
  \includegraphics[width=0.9\textwidth]{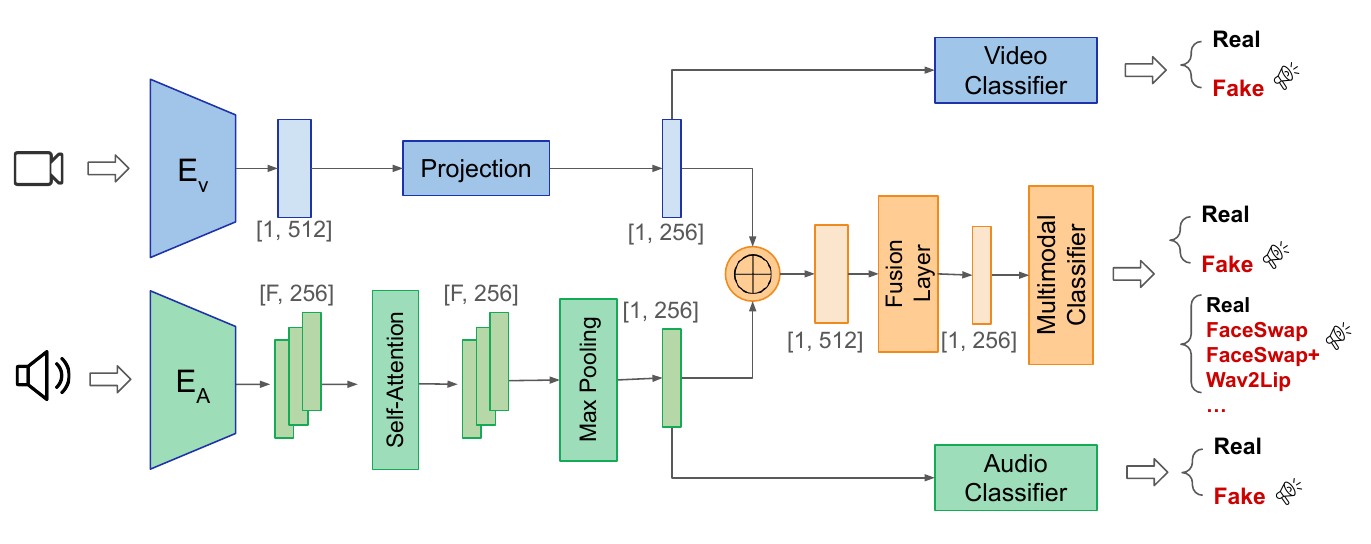}
  \caption{ \textbf{SIMBA} is composed of two encoders, green for audio and blue for the video. A self-attention and a max pooling layer follow the audio encoder. Each encoder has a modality-specific classifier on top. In the fusion, the embedding vectors are concatenated, $\oplus$, followed by a fusion layer and a multimodal classifier. We show both the binary and a multiclass variant. (Best viewed in color.)
  }
  \vspace{-5mm}
  \label{fig:model_architecture}
\end{figure*}

\textbf{Architecture.} \Cref{fig:model_architecture} presents an overview of \ourname. 
R(2D+1)~\cite{r2d1} is the backbone for our video encoder, initialized from Kinetics pretraining~\cite{kinetic_dataset}. The idea behind this model is to separate the classical 3D convolution into a 2D+1D convolution. The first one is used to capture spatial features, and the latter one captures temporal features, leading to an efficient and lightweight model originally for the task of action recognition.
To reduce the dimensionality of the video encoder output and match it to the audio encoder, a projection layer reduces the video feature from $[1, 512]$ to $[1, 256]$.

As an audio encoder, we employ the architecture of the BYOL-A model~\cite{byola}, followed by a self-attention layer and max-pooling.
The self-attention layer is added to enrich the frame-level features along the temporal dimension. The max-pooling layer removes the noisy features and performs dimensionality reduction from $[F, 256]$, where $F$ is the number of audio frames, to $[1, 256]$.

On top of each unimodal branch, a simple binary classification head is added to distinguish between modality-specific real and fake samples. In this way, the model learns to extract the relevant features for the final multimodal classification task. 
The outputs of encoders are concatenated to a vector of dimension $[1, 512]$. The final fusion layer reduces the feature dimension to $[1, 256]$, which is given to the final classification head. 
For the classification, we employ the conventional binary (real/fake) classification, but also investigate a multiclass classification head where each type of manipulation forms its own class (+ the real class). The intuition is that multiclass classification produces a more distinct embedding space during training, which should help generalization. 
During inference, the multiclass predictions are translated back to a binary prediction score by summing all predicted fake probabilities (i.e., all, except for the real class). This sum serves as a final score to describe the ``fakeness'' of the corresponding input sample and is comparable to the binary real/fake prediction score.  
Contrary to the training objective, the evaluation focuses on discovering the presence of a manipulation rather than which specific manipulation it is. %

\textbf{Sampling and Augmentation Strategies.}
To study the robustness and generalizability of our model, we experiment with several sampling and  augmentation strategies during training. Specifically, we study the impact of \emph{temporal jittering} and employ \emph{consecutive frames} vs. \emph{subsampled frames} as our sampling strategies. Temporal jittering describes sampling a clip from the video at a random starting point. This augments the training data and intuitively could increase robustness.
$N$ consecutive frames are sampled as the first sampling strategy. For subsampled frames, $N$ frames are sampled with a step size of $M$. Consecutive frames might provide more information about temporal consistency between two frames, whereas subsampled frames cover a longer temporal window. Concrete hyperparameter choices are provided in~\Cref{sec:experimental_setup}.

\textbf{Loss Functions.}
Our two model variants, binary and multiclass, leverage two distinct losses. The binary classification head is trained with Binary Cross-Entropy Loss (BCE)~\cite{CELoss}. The multiclass classification head is trained with the Cross-Entropy Loss~\cite{CELoss} where individual manipulation types serve as class labels (see~\cref{fig:model_architecture}). The unimodal video and audio classifiers are trained using BCE. The final loss is a sum of the unimodal classifiers together with the multimodal classifier, given as $L_{binary} = BCE_{video} + BCE_{audio} + BCE_{multimodal}$ for \ourname\ binary and $L_{multiclass} = BCE_{video} + BCE_{audio} + CE_{multimodal}$ for \ourname\ multiclass. 

%
%

%

%
%
%
%
%

%

%

%
%
%

%

%

%
%
%

%
%
%
%

%

%
%
%
%
%
%
%
%
%
%
%
%
%
%
%
%
%
%
%
%
%
%
%

\section{Evaluation Protocols}
\label{sec:evaluation_protocols}

\begin{figure*}[t]
    \begin{subfigure}{.48\textwidth}
        \centering
        \includegraphics[width=\linewidth]{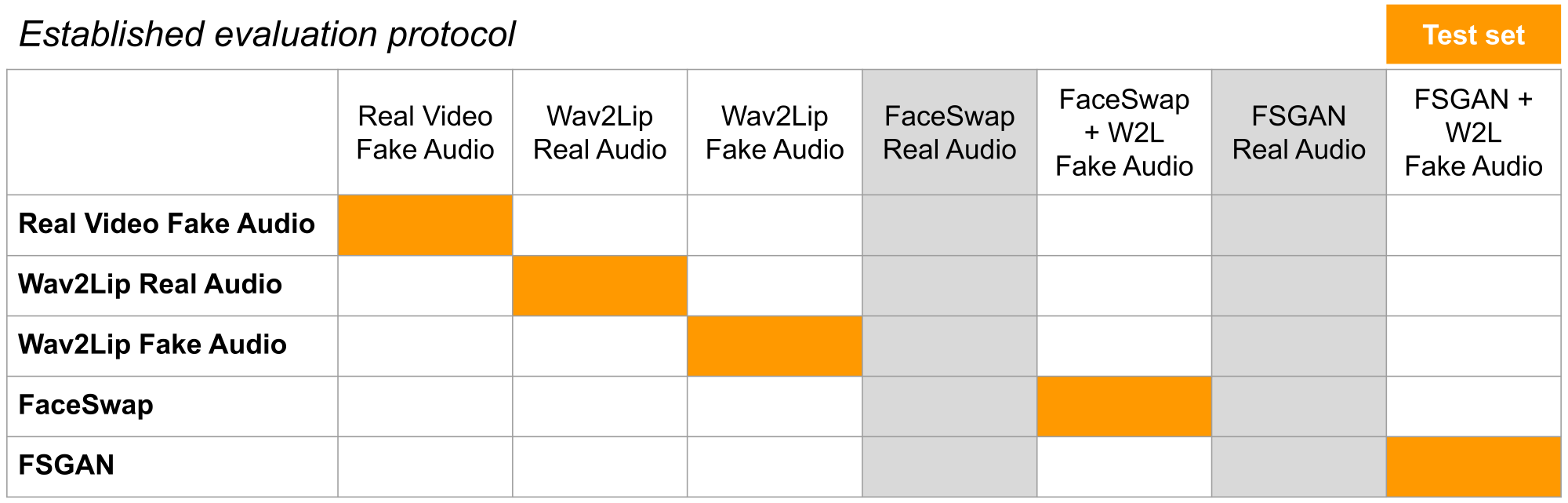}
        \caption{\FAVC\ established evaluation protocol.}
        \label{fig:eval_protocol_favc_established}
    \end{subfigure}%
    \begin{subfigure}{.48\textwidth}
        \centering
        \includegraphics[width=\linewidth]{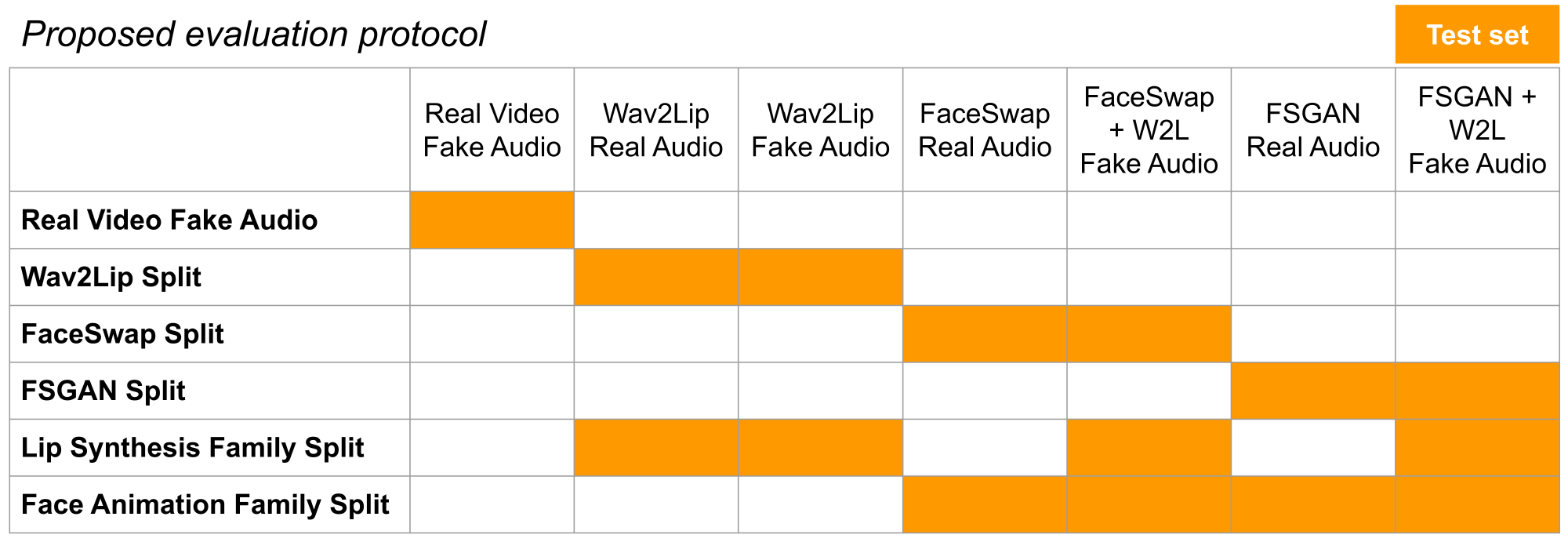}        \caption{\FAVC\ proposed evaluation protocol.}
        \label{fig:eval_protocol_favc_proposed}    \end{subfigure}
   \caption{(a) The established vs. (b) our proposed cross-manipulation generalization evaluation for \FAVC. The top four rows of the proposed evaluation protocol show the different method splits, whereas the last two rows show the family splits Lip Synthesis and Face Animation.}
    \label{fig:eval_protocol_favc}
\end{figure*}

\textbf{Basic Evaluation.} The simplest evaluation scheme is to randomly split the data into training and test (e.g., as 70\%-30\%). In this case, all the manipulations in the test set are also seen in the training set. This scheme typically yields very high performance for supervised approaches, since the models successfully ``capture'' the specific manipulation artifacts observed during training. However, these optimistic results are not representative of the more realistic case of generalizing to unseen scenarios.

\textbf{Leave-one-out (Cross-manipulation) Evaluation.}
As discussed in~\Cref{sec:data_analysis}, our considered datasets contain multiple video manipulations and a single audio manipulation type. Further, there are ``single-manipulation'' and ``multi-manipulation'' (e.g., using \FS\ and \wl\ in combination) fake samples. The common leave-one-out evaluation, where the model is trained on all the manipulations except the one used for testing, for the \FAVC\ dataset (depicted in Fig.~\ref{fig:eval_protocol_favc_established}) was proposed by~\cite{avad}. Originally, it was not intended as a leave-one-out evaluation protocol, yet, follow-up works considered it as one~\cite{AVFF, dimodif}. 
We find several issues with this evaluation protocol. First, some ``single-manipulation'' samples were completely left out, namely \FS\ and \FSGAN. 
Second, the separation of \wl\ real audio and \wl\ fake audio introduces leakage, presenting no generalization task in the visual modality.
Third, some further leakage is introduced with the ``multi-manipulation'' fakes, e.g., in the \wl\ to \FSGAN+\wl\ generalization task, as \wl\ is used in both the manipulations, but they are considered separately.

To allow a more realistic and challenging cross-manipulation evaluation, we propose method and family splits on \FAVC\ (\Cref{fig:eval_protocol_favc_proposed}). The cross-manipulation  (\Cref{fig:eval_protocol_favc_proposed}, top four rows) encompasses one type of video manipulation (``method'') regardless of the audio, e.g., \wl\ real and fake audio form one method, whereas \FS\ and \FS+\wl\ form another method. %
Additionally, we introduce family splits (see~\Cref{fig:eval_protocol_favc_proposed}, last two rows). The Lip Synthesis Family Split includes every modification that has \wl\ (including the combinations with other methods). The Face Animation Family Split consists of all samples that include either \FS\ or \FSGAN. %
This results in a stricter, more challenging, and realistic evaluation setting. %

\begin{wrapfigure}{l}{0.48\textwidth}
\centering
\includegraphics[width=\linewidth]{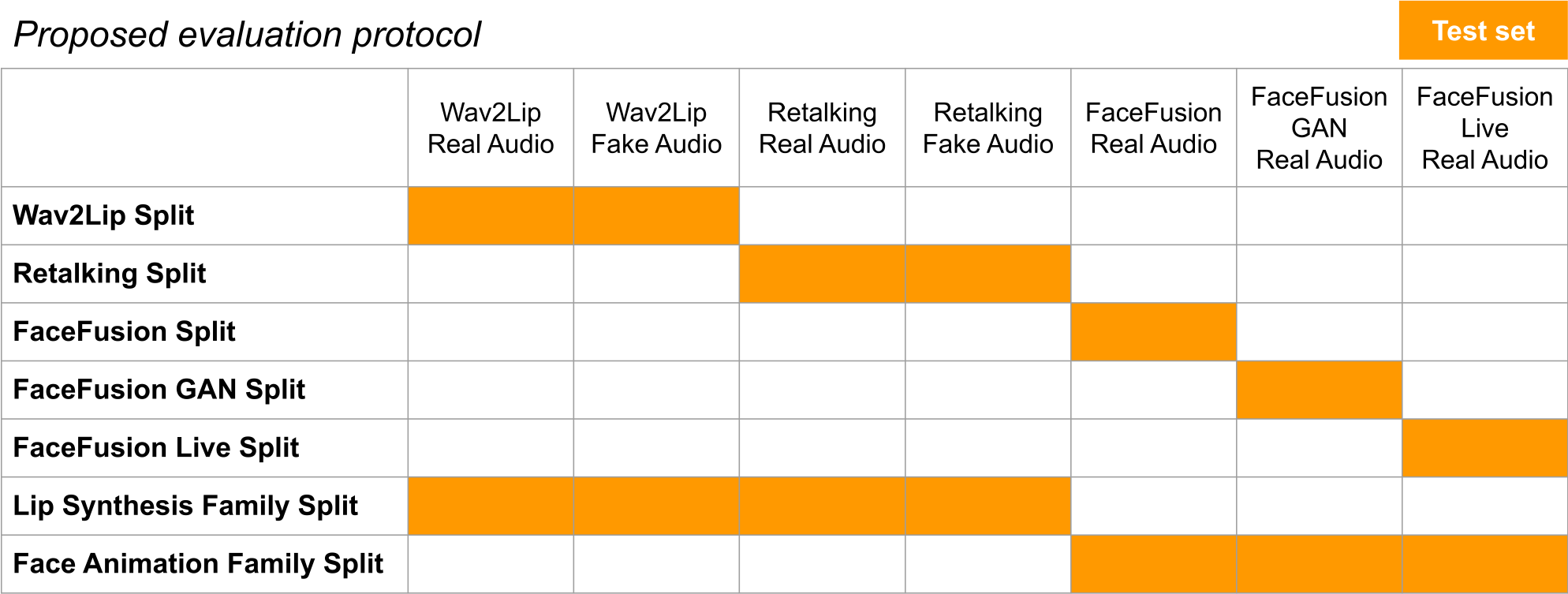}
\caption{The proposed evaluation protocol for \DS. The first five rows show method splits,  the last two rows are the family splits.}
\label{fig:eval_protocol_ds}
\end{wrapfigure}

Similarly, we propose the same evaluation concept for \DS, depicted in~\Cref{fig:eval_protocol_ds}. Again, a method consists of one video manipulation type regardless of the audio. 
This affects only \wl\ and Retalking; Face Fusion, FaceFusion GAN, and FaceFusion Live each form their own split (\cref{fig:eval_protocol_ds}, top 5 rows). 

The family split separates the data into the Lip Synthesis vs. Face Animation Family. The Lip Synthesis Family Split includes the \wl\ and Retalking method. The Face Animation Family Split encompasses FaceFusion, FaceFusion GAN, and FaceFusion Live (\cref{fig:eval_protocol_ds}, bottom 2 rows). In this way, we prevent any leakage of similar artifacts between manipulations and present a harder task for cross-manipulation generalization.

\textbf{Cross-Dataset Evaluation.} Another way to evaluate the generalization abilities of models is to train on one dataset and evaluate on another. Generally, this evaluation is carried out by only considering the \emph{shared manipulation type} between the two datasets (no unseen manipulations), as reported by~\cite{avad, AVFF}. For example, \FAVC\ and KoDF~\cite{kodf} share the \wl\ manipulation, thus the evaluation is limited to the \wl\ vs. real samples. 
Hence, the generalization ability of the models is measured in terms of new domains (different lightning, recording setups, etc). We extend cross-dataset evaluation (between \FAVC\ and \DS) to cover all manipulations, an ultimate challenge to generalize both in terms of domains and manipulation types. 

\section{Experimental Results}

\subsection{Experimental Setup}
\label{sec:experimental_setup}

\textbf{Data Preprocessing.}
We preprocess \DS\ by cropping and resizing the frames to 224x224 pixels around the face regions utilizing the MTCNN~\cite{MTCNN} in a similar way as \FAVC. We set the number of frames to $N=16$ and a stepsize of $M=5$ following AVFF~\cite{AVFF}. The respective audio is converted to a log-mel spectrogram following~\cite{byola}. More details are provided in~\Cref{supp:experimental_setup}.

\textbf{Metrics.} We use average precision (AP) and Area under the curve (AUC). AP is used to measure the precision of the predictions, i.e., the higher the values, the fewer false positives the model predicts. AUC measures how well the model distinguishes between positive and negative classes across all thresholds, where higher is better.%

\textbf{Training Details.} 
We train our \ourname\ models with AdamW with $weight\_decay=0.05$ and $eps=1e-8$, an initial learning rate of $1e-4$. A learning rate scheduler reduces the lr after a plateau of the validation loss for four consecutive epochs. The maximum number of epochs is set to 40, but early stopping intervenes after eight epochs with no validation loss decrease. The batch size is set to 16, distributed among 4xA100 GPUs, and it utilizes $\sim$15 GB of space per GPU. 

\subsection{Benchmarking SIMBA in the basic evaluation}
\label{sec:benchmarkingSIMBA}
We benchmark our \ourname\ model on the standard 70/30$\%$ split of the \FAVC\ alongside LipForensics~\cite{LipForensics}, RealForensics~\cite{RealForensics}, AVoiD-DF~\cite{AVoiD-DF}, AVAD~\cite{avad}, and AVFF~\cite{AVFF}. We show this comparison in Table~\ref{tab:FAVC70_30} to assess \ourname\ using established protocols, although this has to be treated with caution due to the shortcut found in~\cite{CircumventingShortcuts}. \ourname\ performs competitively compared to the SOTA unimodal and multimodal models. Thus, we use it alongside other methods in further analysis.

\begin{wraptable}{r}{0.45\textwidth}
\caption{\FAVC\ performance (in \%). We use the standard 70/30\% split to compare \ourname\ to the SOTA multimodal (AVAD, AVFF, AVoiD-DF) and unimodal (Lip-, RealForensics) methods.}
\label{tab:FAVC70_30}
\centering
\tiny
\begin{tabular}{lcrr} 
    \toprule
    Model                              & Modality & \multicolumn{1}{c}{AUC} & \multicolumn{1}{c}{AP} \\
    LipForensics~\cite{LipForensics}   & V        & 99.81                    & 99.97                  \\
    RealForensics~\cite{RealForensics} & V        & \textbf{99.96}           & \textbf{100.00}        \\ \midrule
    AVoiD-DF \cite{AVoiD-DF}           & AV       & 89.20                    & -                      \\
    AVAD \cite{avad}                   & AV       & 79.16                    & 96.09                  \\
    AVFF \cite{AVFF}                   & AV       & 99.10                    & -                      \\
    SIMBA binary                       & AV       & \textbf{99.91}           & \textbf{99.99}        \\
    SIMBA multiclass                   & AV       & 99.85                    & 99.98        \\
    \bottomrule
\end{tabular}
\end{wraptable}

\subsection{Analyzing the Leading Silence Shortcut}
\label{sec:shortcut_results}
We use \ourname\ as a representative multimodal supervised model to diagnose the silence shortcut in \FAVC\ and \DS. At the same time, we investigate the strategies (temporal jittering and consecutive frames vs. subsampling) introduced in Section~\ref{sec:method} w.r.t. robustness to the shortcut.
Table~\ref{tab:ablation_sampling_strategy} shows the performance of \ourname\ binary and multiclass in a cross-manipulation evaluation on \FAVC. %
To see the impact of the shortcut, we evaluate on untrimmed and trimmed videos (that omit the leading silence). AUC values are given on untrimmed videos, and the difference to the trimmed video performance is given in parentheses. \ourname\ models with consecutive frames and no temporal jittering latch onto the leading silence shortcut, as shown by a significant negative delta when evaluated on trimmed videos. The same finding holds for \ourname\ binary with subsampling, whereas \ourname\ multiclass with subsampling seems to be more robust to the shortcut.

\emph{Introducing the temporal jittering during training reduces the drop in performance between untrimmed and trimmed videos to an almost insignificant delta.}
Comparing consecutive frames vs. subsampling in models trained with temporal jittering shows that subsampling performs slightly better on average. Specifically, subsampling is most beneficial for the realVideo-fakeAudio split. %
Notably, \emph{multiclass \ourname s surpass binary versions on average}.

\begin{table}[ht]
\caption{Silence analysis of \FAVC\ via a cross-manipulation leave-one-out comparison of multiple \ourname\ variants.
Performances are given as AUC on untrimmed videos. The difference to the trimmed video performance is given in parentheses. Significant negative differences ($>10$) are given in \red{red}, for a decrease in performance.}
\label{tab:ablation_sampling_strategy}
\resizebox{\textwidth}{!}{%
\begin{tabular}{@{}lrrrrr@{}}
\toprule
\multicolumn{1}{c}{}        & \multicolumn{1}{c}{AVG} & \multicolumn{1}{c}{\begin{tabular}[c]{@{}c@{}}Wav2Lip\\ Split\end{tabular}} & \multicolumn{1}{c}{\begin{tabular}[c]{@{}c@{}}realVideo\\ fakeAudio\\ Split\end{tabular}} & \multicolumn{1}{c}{FSGAN Split} & \multicolumn{1}{c}{\begin{tabular}[c]{@{}c@{}}FaceSwap\\ Split\end{tabular}} \\ \midrule
Binary consecutive         & 90.39 (\red{-10.89})          & 100.00 (-4.08)                                                              & 74.26 (\red{-35.59})                                                                            & 100.00 (-0.77)                  & 87.30 (-3.11)                                                                \\
Multiclass consecutive     & 95.24 (\red{-15.76})          & 99.25 (-9.55)                                                               & 99.98 (\red{-46.29})                                                                            & 99.98 (-3.69)                   & 81.74 (-3.49)                                                                \\ \midrule
Binary consecutive jit     & 87.22 (-0.81)           & 100.00 (+0.00)                                                              & 67.09 (-4.14)                                                                             & 99.98 (+0.01)                   & 81.79 (+0.89)                                                                \\
Multiclass consecutive jit & 93.91 (-0.17)           & 99.37 (+0.37)                                                               & 93.87 (-1.25)                                                                             & 100.00 (+0.00)                  & 82.39 (+0.21)                                                                \\ \midrule
Binary subsampling         & 95.15 (\red{-12.51})          & 100.00 (+0.00)                                                              & 92.65 (\red{-45.06})                                                                            & 100.00 (-4.03)                  & 87.95 (-0.93)                                                                \\
Multiclass subsampling     & 94.42 (-1.16)           & 99.99 (-0.41)                                                               & 99.98 (-3.60)                                                                             & 100.00 (-0.01)                  & 77.71 (-0.62)                                                                \\ \midrule
Binary subsampling jit     & 89.48 (-0.54)           & 99.98 (+0.00)                                                               & 81.00 (-1.76)                                                                             & 100.00 (+0.00)                  & 76.93 (-0.40)                                                                \\
Multiclass subsampling jit & 95.34 (-0.34)           & 99.41 (-0.01)                                                               & 99.32 (-0.86)                                                                             & 100.00 (+0.00)                  & 82.61 (-0.50)                                                               
\end{tabular}%
}
\end{table}

We also diagnose a possible shortcut on \DS\ in Table~\ref{tab:ablation_ds_sampling_strategy}, together with the consecutive frames vs. subsampling dimension. Note that both \ourname\ models trained on consecutive frames without temporal jittering show a significant negative delta when evaluated on untrimmed vs trimmed videos. This negative delta is especially present for both models on the \wl\ and Retalking split, suggesting that \emph{there is indeed a leading shortcut in these splits}. This supports the observation of the imbalance in the first bins in Figure~\ref{fig:ds_leadingsilence}, where \wl\ and Retalking dominate over the other manipulations and Real videos. Besides, \ourname\ binary has an unexpected drop on FaceFusion GAN and Live (top row), likely a side-effect of the learned shortcut.

Again, applying temporal jittering to \ourname\ models reduces these negative deltas significantly. \emph{This simple and intuitive technique can be easily incorporated in most other multimodal models.}
Training with subsampling achieves slightly higher results on average than with consecutive frames, although consecutive frames seem to have a large impact on FaceFusion GAN. Surprisingly, after trimming the silence, the AUC on these samples improves significantly. The difference between \ourname\ binary and multiclass is overall not as clear as on \FAVC.

\begin{table}[ht]
\caption{Silence analysis of \DS\ via a cross-manipulation leave-one-out comparison of multiple \ourname\ variants. Performances are given as AUC on untrimmed videos. The difference to the trimmed video performance is given in parentheses. Significant differences ($>3$) are given in \red{red}/\green{green} for a decrease/increase in performance.}
\label{tab:ablation_ds_sampling_strategy}
\resizebox{\textwidth}{!}{%
\begin{tabular}{@{}lrrrrrr@{}}
\toprule
                            & \multicolumn{1}{l}{AVG} & \multicolumn{1}{l}{\begin{tabular}[c]{@{}l@{}}Wav2Lip\\ Split\end{tabular}} & \multicolumn{1}{l}{\begin{tabular}[c]{@{}l@{}}Retalking\\ Split\end{tabular}} & \multicolumn{1}{l}{\begin{tabular}[c]{@{}l@{}}FaceFusion\\ Split\end{tabular}} & \multicolumn{1}{l}{\begin{tabular}[c]{@{}l@{}}FaceFusion\\ GAN Split\end{tabular}} & \multicolumn{1}{l}{\begin{tabular}[c]{@{}l@{}}FaceFusion\\ Live Split\end{tabular}} \\ \midrule
Binary consecutive         & 91.89 (\red{-5.27})           & 99.43 (\red{-5.15})                                                               & 97.42 (\red{-8.23})                                                                 & 92.23 (+0.00)                                                                  & 72.48 (\red{-8.90})                                                                      & 97.87 (\red{-4.08})                                                                       \\
Multiclass consecutive     & 89.97 (\red{-3.00})           & 98.60 (\red{-4.16})                                                               & 94.62 (\red{-8.67})                                                                 & 90.31 (-2.93)                                                                  & 68.07 (+0.26)                                                                      & 98.23 (+0.48)                                                                       \\ \midrule
Binary consecutive jit     & 89.62 (+2.44)           & 97.86 (+0.38)                                                               & 87.91 (-2.07)                                                                 & 93.81 (+2.90)                                                                  & 69.63 (\green{+10.03})                                                                     & 98.89 (+0.95)                                                                       \\
Multiclass consecutive jit & 92.95 (+2.22)           & 99.39 (+0.20)                                                               & 91.24 (+2.09)                                                                 & 88.12 (+2.56)                                                                  & 86.06 (\green{+6.20})                                                                      & 99.95 (+0.05)                                                                       \\ \midrule
Binary subsampling jit     & 93.59 (+0.56)           & 99.10 (+0.25)                                                               & 93.80 (+0.36)                                                                 & 96.51 (+0.37)                                                                  & 78.64 (+1.82)                                                                      & 99.89 (+0.01)                                                                       \\
Multiclass subsampling jit & 93.06 (+0.13)           & 99.51 (+0.07)                                                               & 92.44 (-0.26)                                                                 & 95.26 (-0.01)                                                                  & 78.41 (+0.85)                                                                      & 99.70 (+0.00)                                                                       \\ \bottomrule
\end{tabular}%
}
\end{table}

\subsection{Cross-manipulation Generalization Evaluation}
\label{sec:cross_manipulation_results}
Next, we evaluate models that are robust to the leading silence shortcut using our newly proposed evaluation protocols. Our considered models are Lip-, RealForensics~\cite{LipForensics, RealForensics}, AVAD~\cite{avad}, and our multimodal supervised baseline \ourname\ trained with subsampling and temporal jittering (abbr. ``jit''). Notice that AVAD is not evaluated in the conventional leave-one-out scenario as it is trained in a self-supervised fashion on LRS2~\cite{LRS2}, i.e., it generalizes to all manipulations simultaneously. 

\begin{figure*}[t]
    \begin{subfigure}{.5\textwidth}
        \centering
        \includegraphics[width=\textwidth]{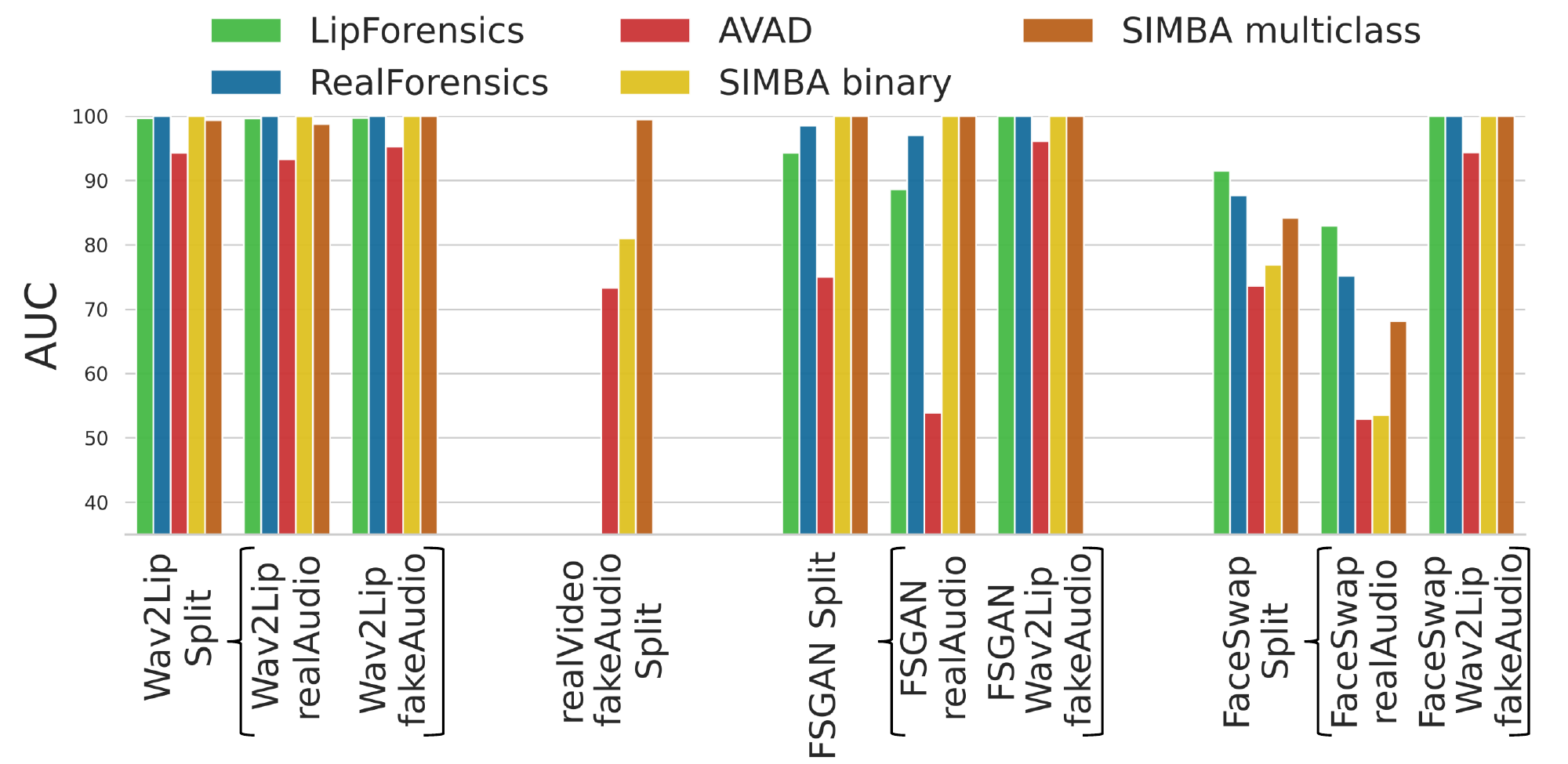}
        \caption{\FAVC}
        \label{fig:method_split_favc}
    \end{subfigure}%
    \begin{subfigure}{.5\textwidth}
        \centering
        \includegraphics[width=\textwidth]{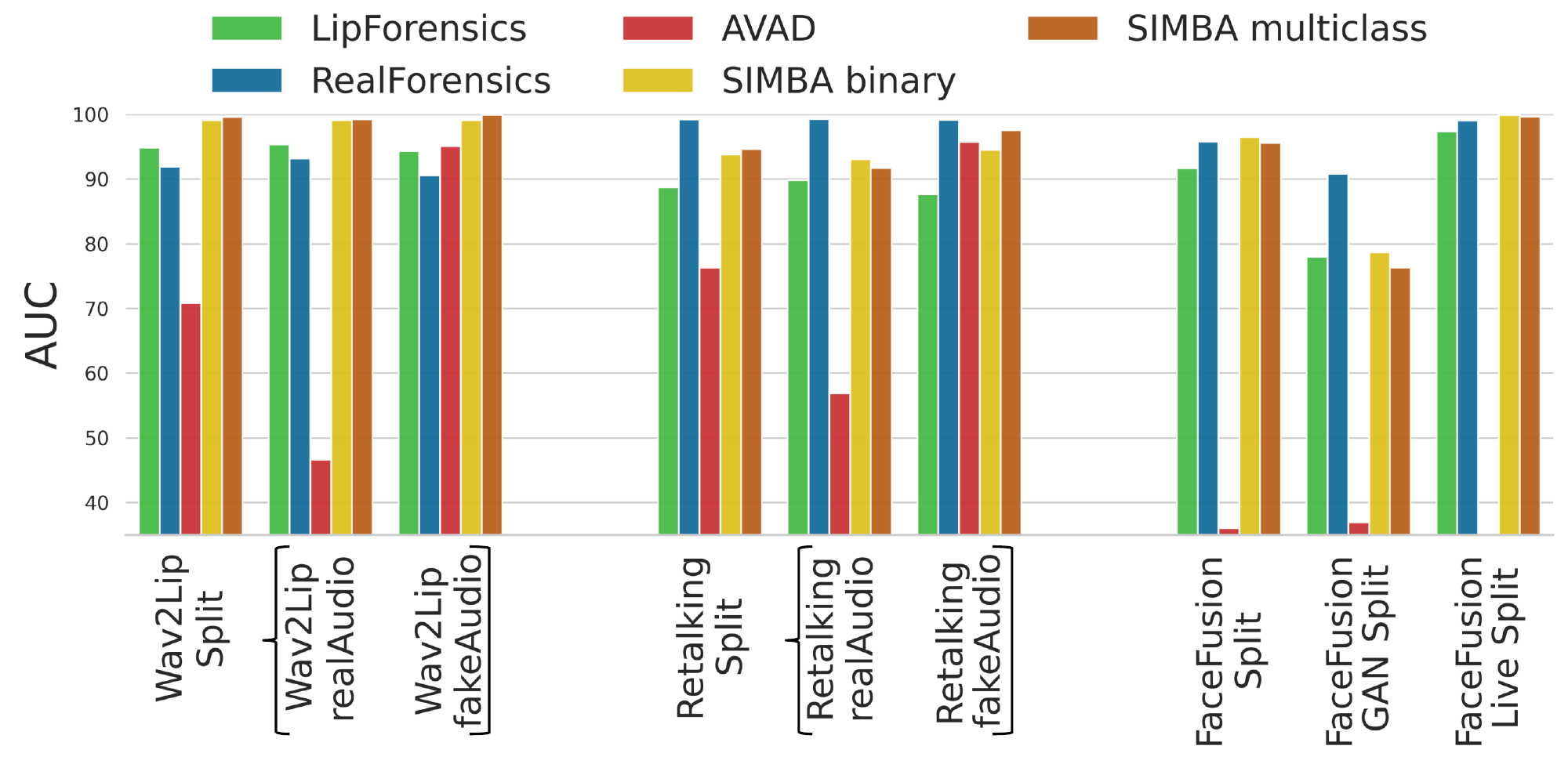}
        \caption{\DS}
        \label{fig:method_split_ds}
    \end{subfigure}
    \caption{Cross-manipulation comparison using the proposed methods splits (as AUC).}
    \label{fig:method_split}
\end{figure*}

Figure~\ref{fig:method_split_favc} shows results using our method splits on \FAVC. All models show high performance on \wl, suggesting that \emph{this is the easiest split to generalize to}. The same holds for the \FSGAN\ split for the supervised models. Hereby, we ``reveal'' the performance on the \FSGAN\ realAudio subset, which was hidden in the old evaluation protocol. \emph{The other previously hidden subset is \FS\ realAudio, which seems to be the hardest subset to generalize to for all models.} Here, the unimodal models achieve higher performance than the multimodal models. We hypothesize that the latter may be slightly more tailored for lip-syncing rather than face animation manipulations.

Results for the method splits on \DS\ are shown in Figure~\ref{fig:method_split_ds}. 
Here, almost all supervised models result in performances $>90\%$ AUC on the lip synthesis-related splits. All supervised models perform competitively on the different FaceFusion splits. Our method breakdown reveals that FaceFusion GAN split is the hardest to generalize to, whereas FaceFusion Live is the easiest. \emph{The self-supervised AVAD only shows competitive performance when the audio is fake.} 

Results for the family splits for both datasets are provided in~\Cref{supp:family_split_results}.

\subsection{Cross-Dataset Generalization Evaluation}
\label{sec:cross_dataset_results}
\begin{figure}[t]
    \begin{subfigure}{.5\textwidth}
        \centering
        \includegraphics[width=\textwidth]{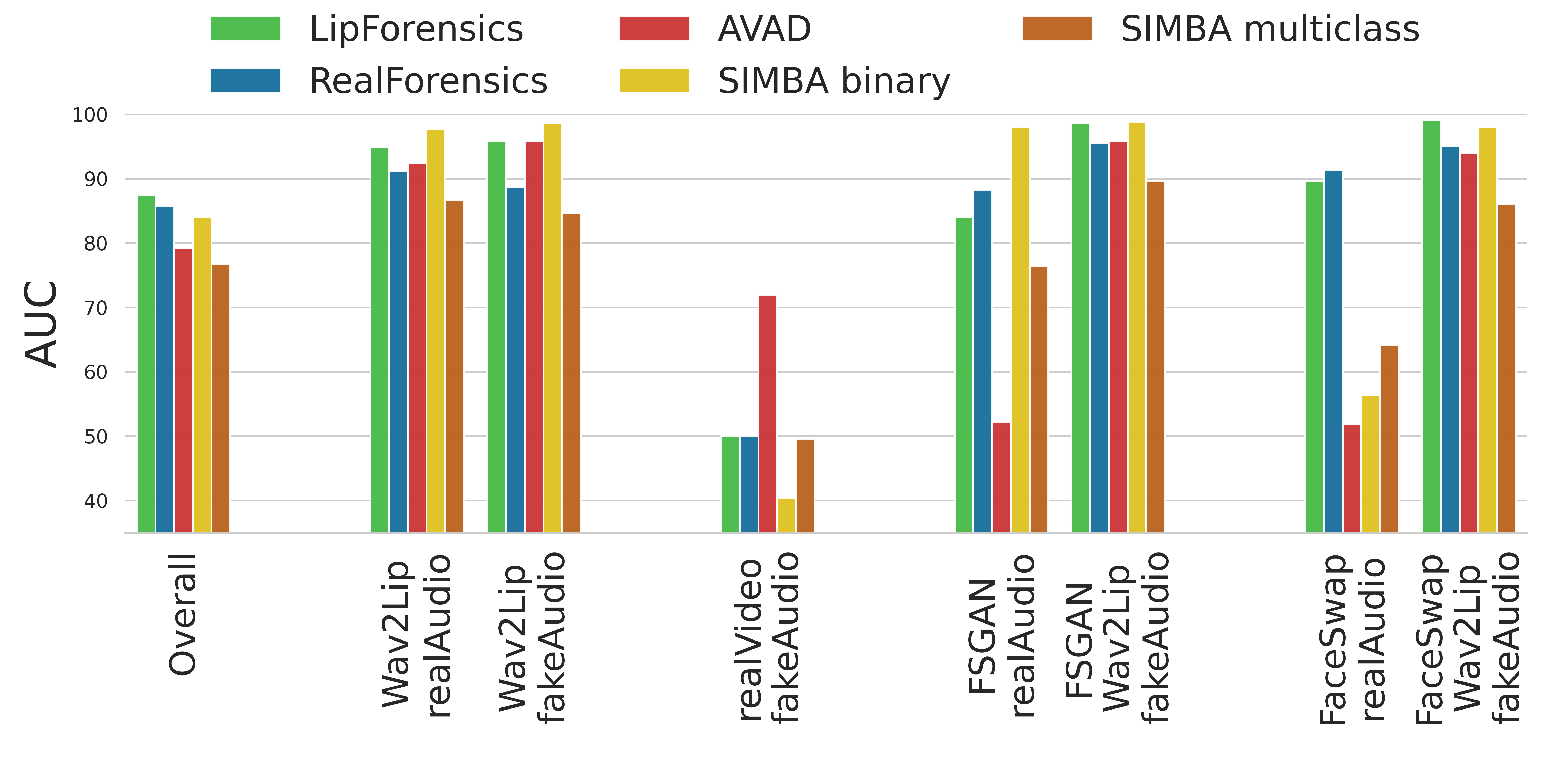}
        \caption{\DS\ to \FAVC\ generalization}
        \label{fig:ds2favc}
    \end{subfigure}%
    \begin{subfigure}{.5\textwidth}
        \centering
        \includegraphics[width=\textwidth]{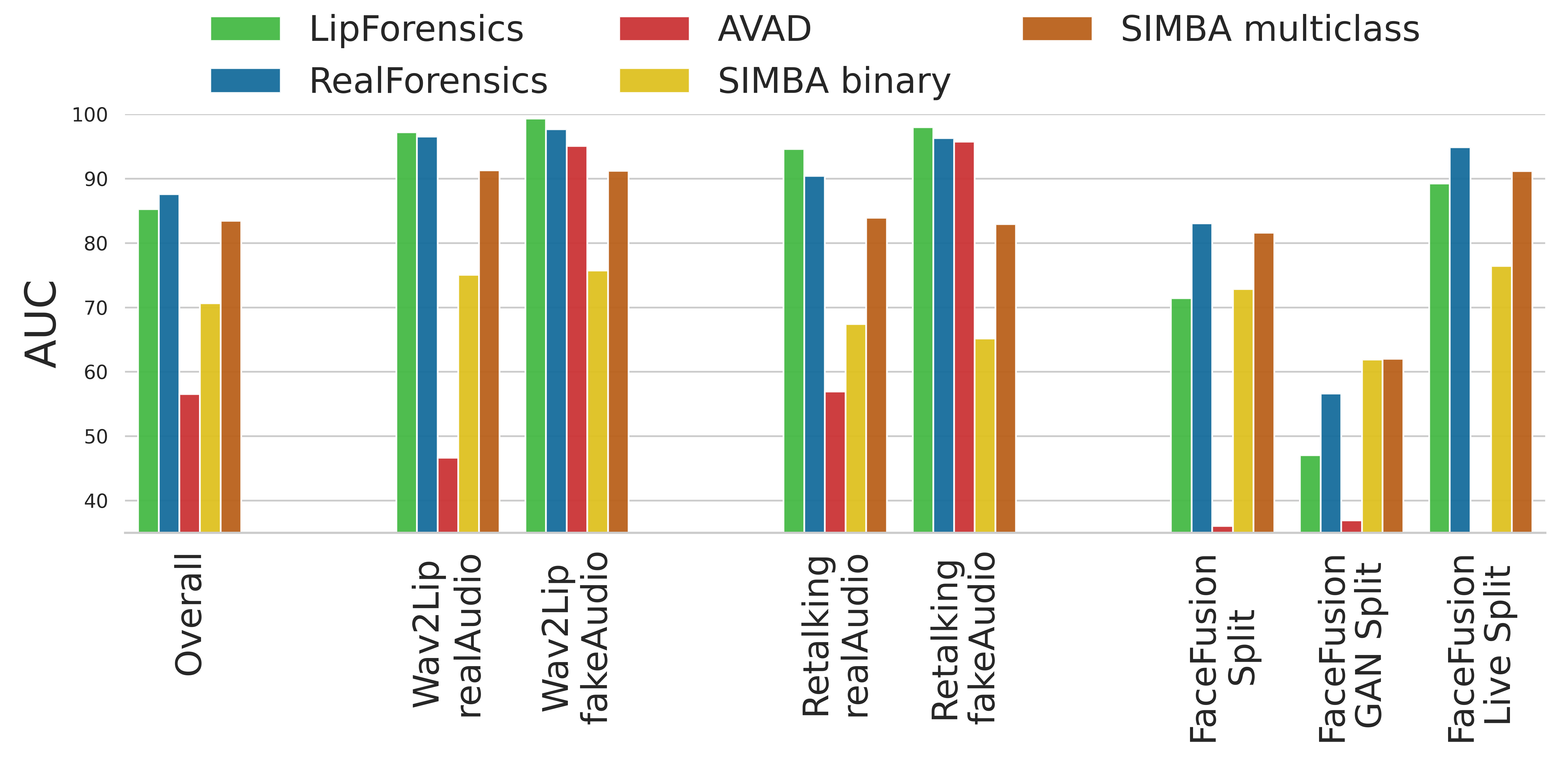}
        \caption{\FAVC\ to \DS\ generalization}
        \label{fig:favc2ds}
    \end{subfigure}
    \caption{Cross-dataset evaluation across all manipulations (as AUC)}
    \label{fig:cross-dataset}
\end{figure}

Figure~\ref{fig:cross-dataset} shows performance of Lip-, RealForensics, AVAD, and our \ourname\ models in a cross-dataset generalization task, meaning each model was trained on all manipulations of the training dataset and evaluated on all manipulations of the test dataset. Recall that AVAD generalizes from LRS2 to the respective test dataset. As this is a multimodal task, we set the AUCs of the unimodal models to $50\%$ for realVideo-fakeAudio as they can not handle this split.

Surprisingly, \emph{Lip- and RealForensics show very strong generalization performances in both directions}, beating all the other models on average. \ourname\ models struggle when generalizing from \DS\ to \FAVC\ on the realVideo-fakeAudio split, where AVAD's focus on temporal alignment is more beneficial. 
Interestingly, the results of AVAD on \wl\ realAudio are significantly lower on \DS\ than on \FAVC. This hints at a larger domain gap between its training dataset (LRS2) and \DS\ compared to the gap between LRS2 and \FAVC. As \DS\ seems to be more out-of-distribution compared to established datasets, it supports our case for using this more recent dataset.

Generalizing from \FAVC\ to \DS\ is overall a more challenging task, resulting in lower numbers on average.
It is interesting that video-only approaches surpass multimodal models on average, despite both datasets including audio manipulations. This underscores that the \emph{datasets are somewhat skewed towards video modality}.
Our evaluation shows that \emph{jointly generalizing to a new manipulation and a new dataset is challenging yet not impossible for SOTA methods}.

\section{Discussion, Limitations, and Broader Impact}
\label{sec:discussion}

To sum up, in this work, we contributed to the multimodal \DF\ detection by diagnosing the benchmarking issues along the axes of \emph{datasets}, \emph{methods}, and \emph{evaluation protocols}. We showed that the recent \DS\ dataset is a suitable benchmark with room for improvement. We presented a baseline method \ourname, and showed how temporal jittering augmentation scheme leads to robustness to the shortcut issue. We disclosed issues in the existing \FAVC\ evaluation protocol, and offered new protocols for both the \DS\ and \FAVC\ datasets. 

Our work has limitations, too, which are partly due to the core issues we aim to address: limited implementation availability of prior approaches makes empirical comparison to these methods challenging. Similarly, various deficiencies of the prior datasets restrict further benchmarking due to the data being unavailable or incompatible with our analysis. We hope that our work overall has a positive societal impact, as we aim to advance the important topic of multimodal DeepFake detection to counter the spread of misinformation. As with any technology, there is some potential for misuse; we can not rule out that our models learn biased representations since the training datasets potentially may contain some biases. We recommend caution when using the models.

Overall, we emphasize the need for new diverse datasets, placing priority on reproducibility and standardized benchmarking, to enable further progress in audio-video \DF\ detection.

\paragraph{Acknowledgement} For compute, we gratefully acknowledge support from the hessian.AI Service Center (funded by the Federal Ministry of Education and Research, BMBF, grant no. 01IS22091) and the hessian.AI Innovation Lab (funded by the Hessian Ministry for Digital Strategy and Innovation, grant no. S-DIW04/0013/003).

\medskip
{
    \bibliographystyle{plainnat}
    \bibliography{biblio}
}

\appendix

\clearpage

\begin{center}
    {{\includegraphics[scale=0.1]{figures/doctor-logo1.png}}\Large\textbf{DeepFake Doctor: \\ Diagnosing and Treating Audio-Video Fake Detection}} \\
    \large\textbf{Supplementary Material}\\[1.5em]
\end{center}

\setcounter{page}{1}

\maketitle

The supplementary materials are organized as follows:

\Cref{supp:protocol} presents the details on how the datasets, \DS\ and \FAVC, were split into training, validation, and test sets. 

\Cref{supp:multiclass} provides information on how the metrics of AP and AUC are computed for \ourname\ in the multiclass scenario. 

\Cref{supp:experimental_setup} introduces more details on our experimental setup. 

\Cref{supp:validating_simba} validates \ourname\ by comparing it to SOTA models on \FAVC\ using the established leave-one-out evaluation protocols.

\Cref{supp:family_split_results} shows and discusses the performance of SOTA and \ourname\ models using our family splits for \FAVC\ and \DS.

\Cref{supp:ablation_eval_sampling} offers an ablation study on different sampling strategies during inference. 

\Cref{supp:embeddings} offers a graphic representation of the embedding spaces for both the \ourname\ binary and multiclass. 

\section{Details on the Datasets}
\label{supp:protocol}

This section introduces some details on how \DS~\cite{deepspeak_dataset} and \FAVC~\cite{khalid2021fakeavceleb} are split into training, validation, and test sets, and shows some examples of per-manipulation samples.  We introduce a validation set for both datasets, which is used for learning rate scheduling and early stopping. Additionally, the newly introduced set was used for initial architecture decisions and hyperparameter selection. Method and Family Splits are created by leaving out the corresponding manipulations from the training and validation split and then evaluating only these manipulations in the test set. In other words, there is \emph{no leakage} between training plus validation, and test. 

\begin{figure}[ht]
	\centering
    \begin{subfigure}[b]{0.48\linewidth}
        \centering
        \includegraphics[width=\textwidth]{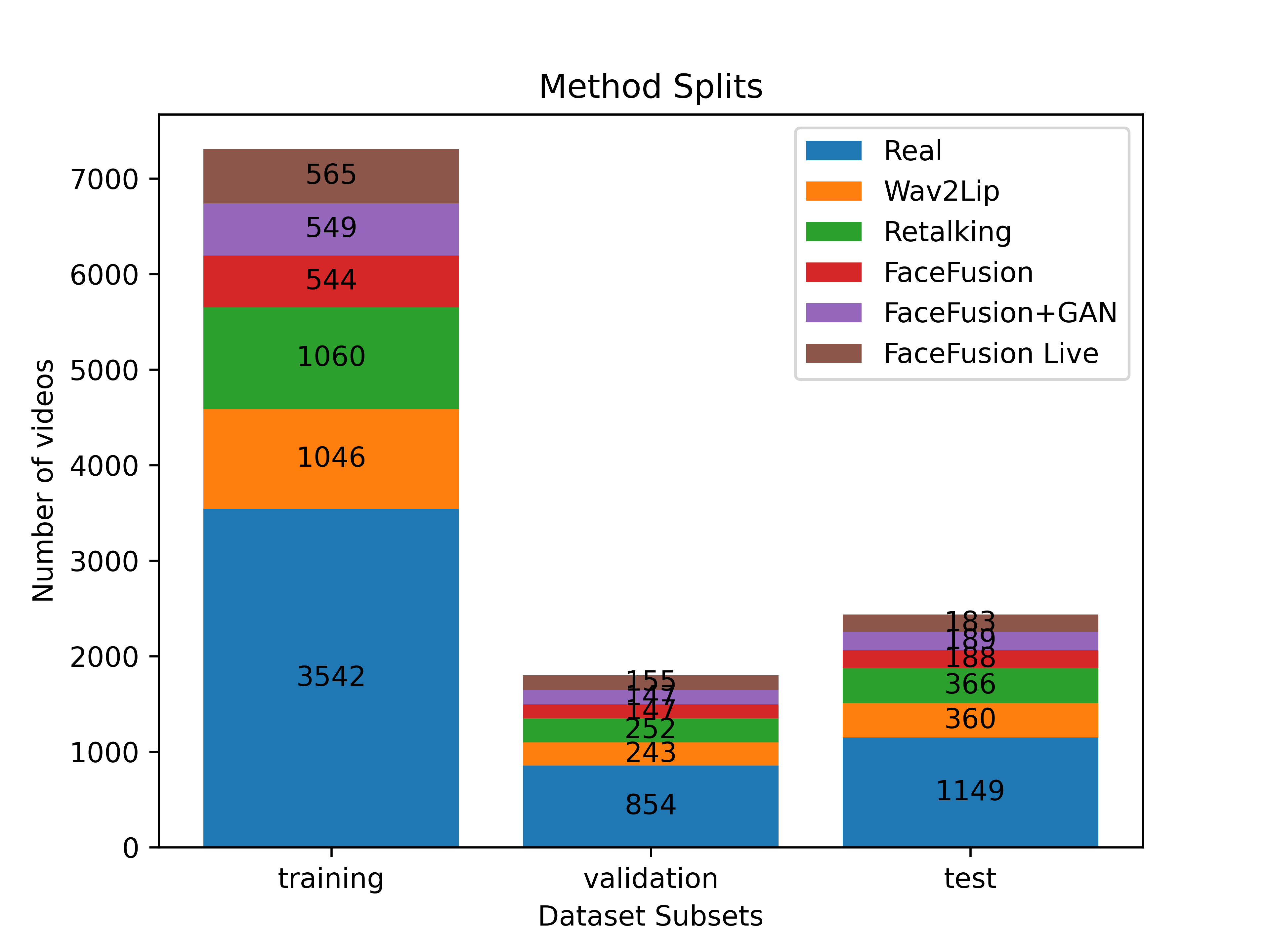}
        \caption{\DS}
        \label{fig:DS_splits}
    \end{subfigure}
    \begin{subfigure}[b]{0.48\linewidth}
        \centering
        \includegraphics[width=\textwidth]{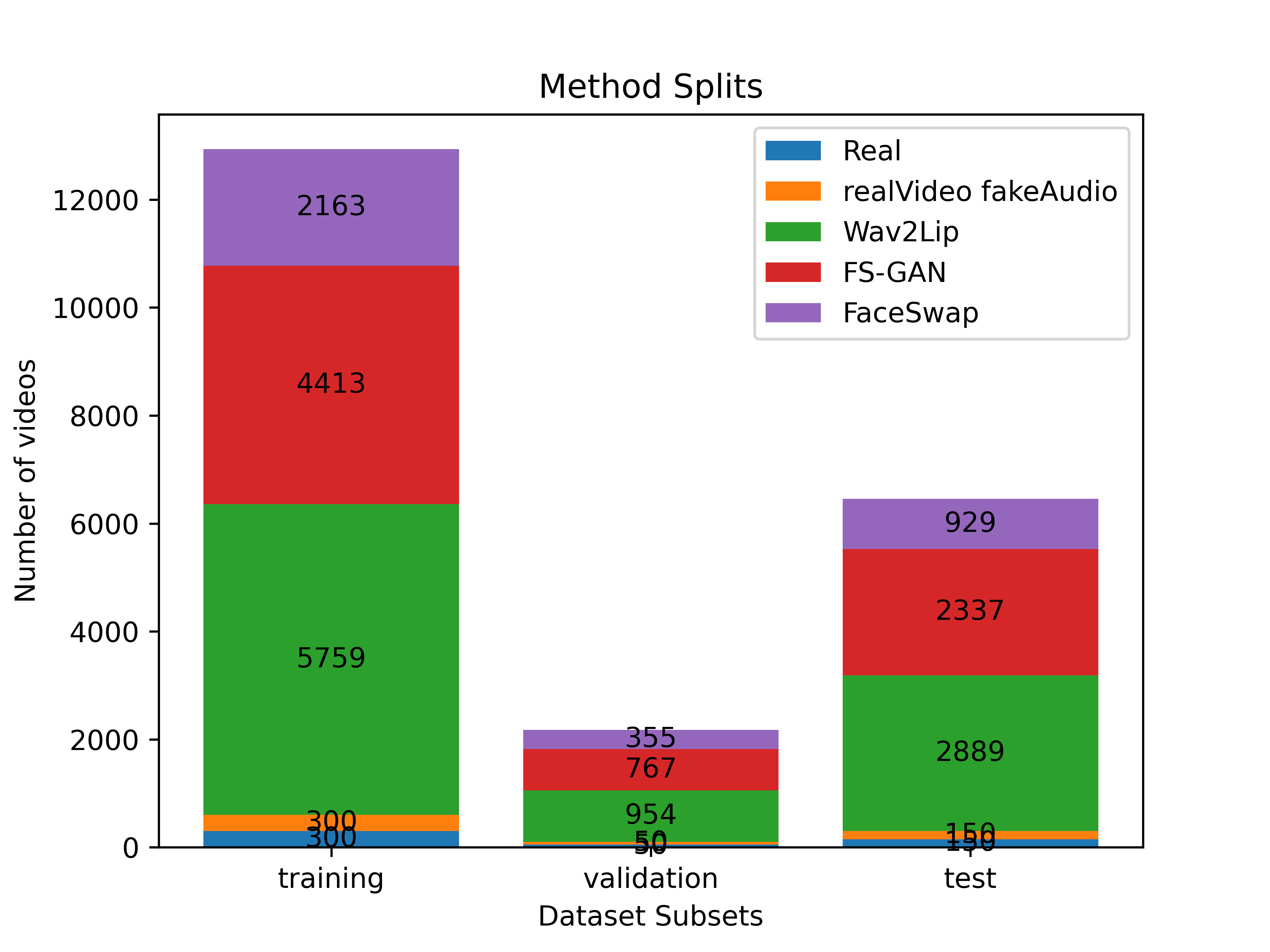}
        \caption{\FAVC}
        \label{fig:FAVC_splits}
    \end{subfigure}    
	\caption{Number of videos for each method split in training, validation, and test for \FAVC and \DS.}%
	\label{fig:method_split_distribution}
\end{figure}

\begin{figure*}[ht]
    \centering
    \includegraphics[width=0.99\linewidth]{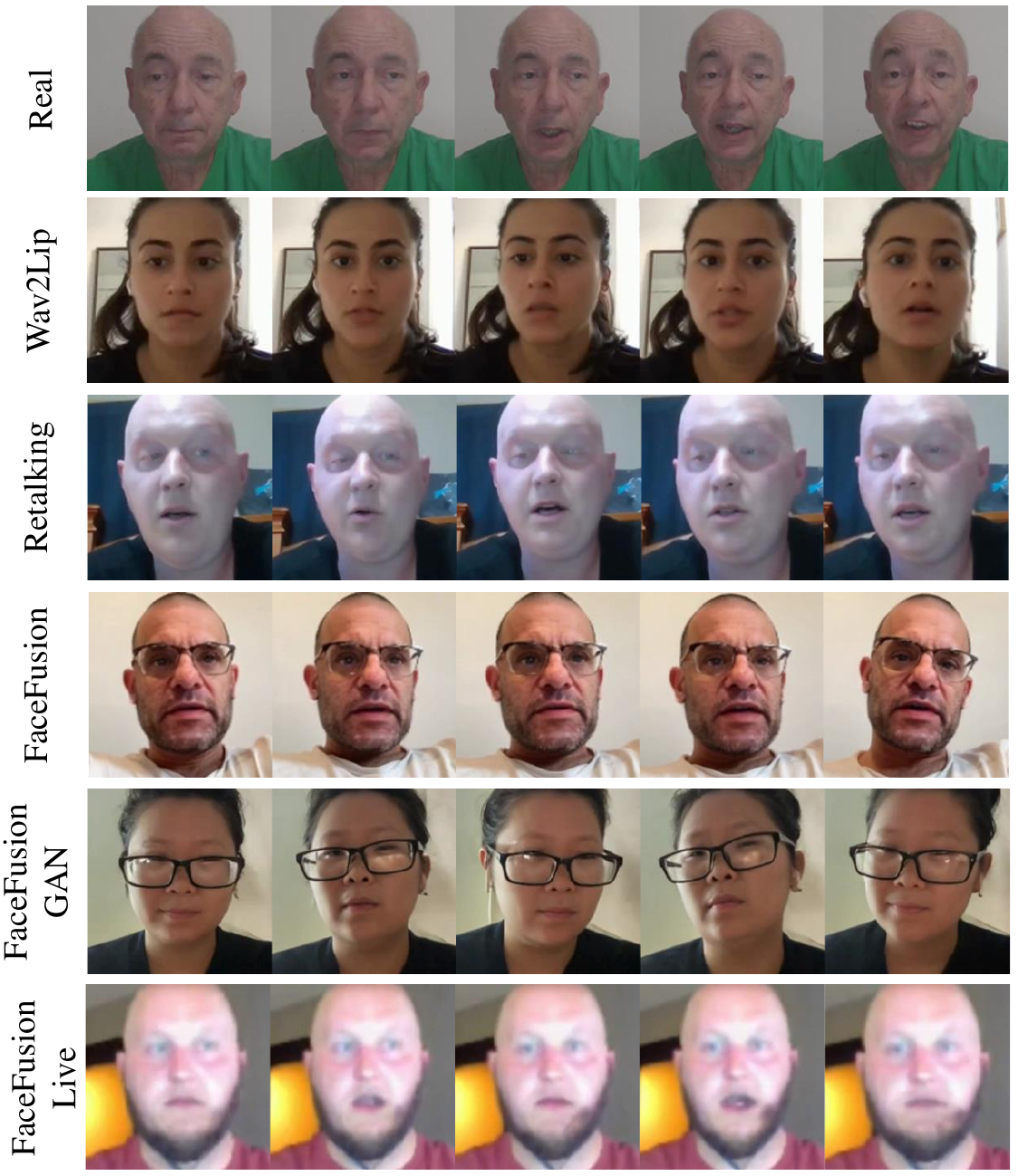}
    \caption{Examples of the video manipulation types in \DS.}
    \label{fig:DS_examples}
\end{figure*}

\paragraph{\DS} We leave the test set provided by \DS~\cite{deepspeak_dataset} unchanged, however, we use $20\%$ of the training data as a validation set. Method and Family Splits are constructed in the same way as above: E.g., leaving out FaceFusion in the training and validation set and evaluating on real vs. FaceFusion samples in the test set yields the FaceFusion Split (see~\cref{fig:DS_examples} for some examples of the samples). In total, the training set encompasses 7,306 samples, the validation set 1,798, and the test set 2,435. Figure~\ref{fig:DS_splits} provides a detailed breakdown of the number of samples per method split in each set. 

\begin{figure*}[ht]
    \centering
    \includegraphics[width=0.99\linewidth]{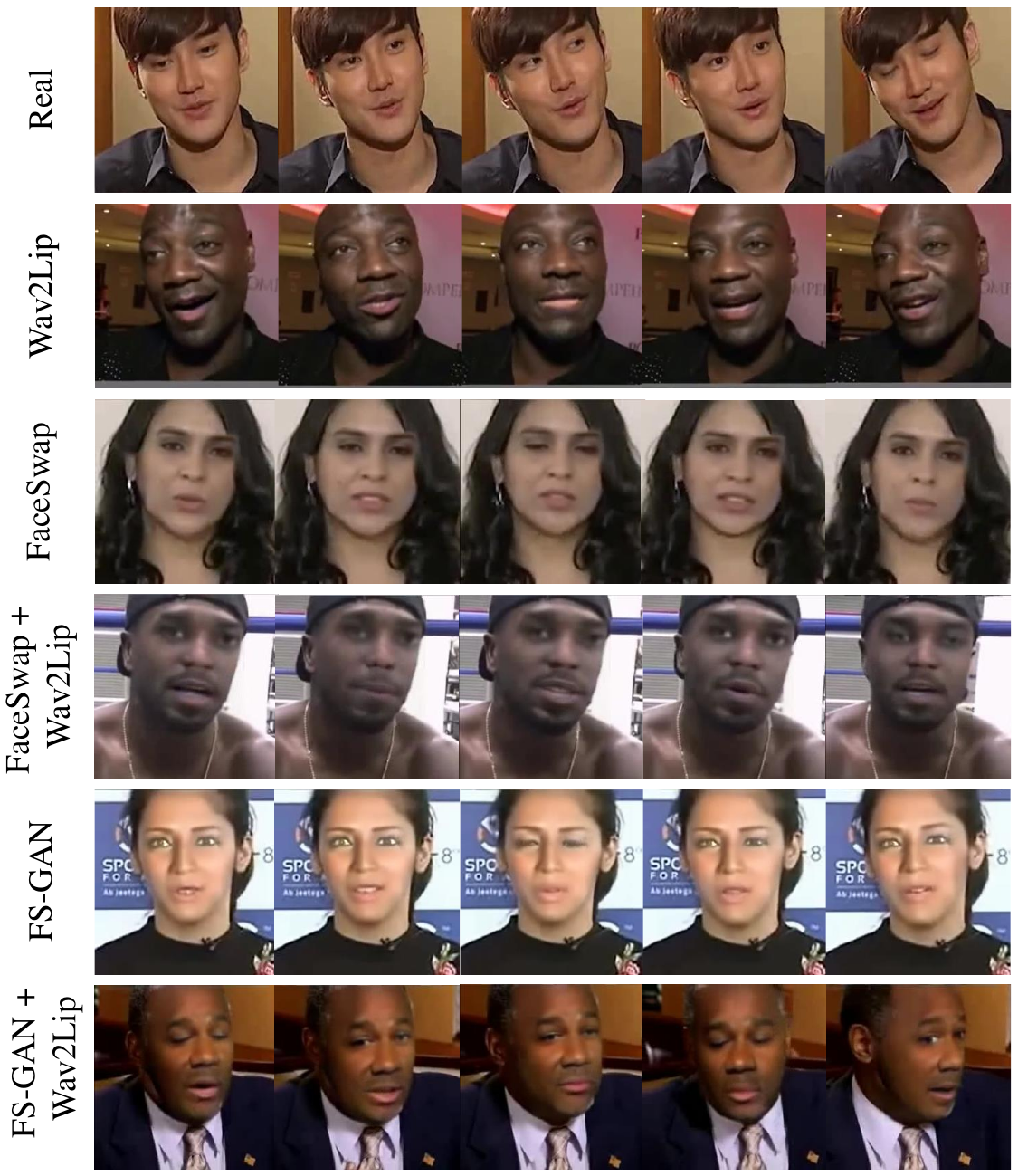}
    \caption{Examples of the video manipulation types in \FAVC.}
    \label{fig:FAVC_examples}
\end{figure*}

\paragraph{\FAVC} We split the \FAVC~\cite{khalid2021fakeavceleb} into $60\%$ training, $10\%$ validation, and $30\%$ test set based on the identity of the person, obtained from the provided annotations. (For the fakes, we use the source identity of a fake.) Thus, if id-0 is selected as a training identity, the real sample of id-0 and all the fake samples with id-0 as the source identity are included in the training set. 
Further, the Method Splits are created by removing the corresponding manipulation method from the training and validation set, and evaluating on the respective (held-out) manipulations in the test set. For example, the \wlS\ is created by removing the \wl\ method from the training and validation set and evaluating on the real vs. all \wl\ samples present in the test set (see~\cref{fig:FAVC_examples} for some examples of the samples). 
The standard split (nothing is held-out) into training, validation, and test set contain $12,935$, $2,176$, and $6,455$ samples, respectively. The resulting number of samples per method split in each set can be found in Figure~\ref{fig:FAVC_splits}.

Note that the two datasets are rather different from each other in terms of the amount of real samples: a lot more in the \DS\ case, which is a more realistic scenario.

\section{SIMBA Multiclass Evaluation}
\label{supp:multiclass}
\emph{SIMBA multiclass} predicts a score distribution over all the available training classes. As the final decision we are interested in is a binary decision on whether the test input is real or fake, the multiclass score distribution is post-processed to obtain a single confidence score. For that, all the scores of fake classes from the softmax distribution are summed for each sample: %
\begin{equation}
    score = \sum^{C}_{c=2} conf\ score_c,
    \label{eq:score_summation}
\end{equation}
where $C$ is the number of classes and $c == 1$ is the real class. This confidence score is used for AP and AUC calculation. 
A straightforward approach would be to apply a default threshold of 0.5 for a binary ``real''/``fake''-decision, which then can be used to calculate accuracy. 
However, we found that using the $argmax$ operation on the multiclass prediction to determine the most similar (training) class works better. The required binary decision is then obtained by treating every predicted class that is not the real class as fake. %

\section{Detailed Experimental Setup}
\label{supp:experimental_setup}
\paragraph{DeepSpeak Preprocessing} We preprocess \DS~\cite{deepspeak_dataset} by cropping and resizing the frames to 224x224 pixels around the face regions utilizing the MTCNN~\cite{MTCNN} in a similar way as \FAVC~\cite{khalid2021fakeavceleb}.
\paragraph{Training Details} During training, we selected from the video a number of frames equal to $N=16$ and a stepsize of $M=5$ following AVFF~\cite{AVFF}. Padding is applied where necessary. For audio, we keep the sampling rate of the original audio samples equal to $16000 Hz$ for \FAVC and to $48000Hz$ for \DS. Additionally, the respective audio is converted to a log-mel spectrogram, with a $n_{ftt}=321$ and $n_{mels}=64$, following BYOL-A settings~\cite{byola}. The audio is then normalized by computing the $log$ of the spectrogram and normalizing this value with the mean and standard deviation, computed among the audio samples of the dataset. 

Regarding the hyperparameters, the learning rate is $1e-4$ and we use the \emph{ReduceLROnPlateau}, with a patience of $4$ epochs. The total number of epochs is $40$, with an Early Stopping value of $8$ epochs. 

The self-attention layer after the audio encoder is defined with a number of layers equal to $2$ and of attention heads equal to $8$.

\section{Validating SIMBA}
\label{supp:validating_simba}
Besides validating our \ourname\ models on the 70/30 split (Tab.~\ref{tab:FAVC70_30}), we benchmark \ourname\ using the former ``leave-one-out'' evaluation schema (Fig.~\ref{fig:eval_protocol_favc_established}) on \FAVC\ in Table~\ref{tab:FAVC_established_loo}. 
\ourname\ results are comparable, or even better, than current SOTA models. Especially strong is \ourname\ on the rV-fA split where it reaches almost 99\% AUC, surpassing the other two multimodal models and underlining the strengths of \ourname s audio encoder. This supports our use of \ourname\ as a supervised multimodal analysis tool.

\begin{table*}[th]
\caption{Leave-one-out comparison between \ourname\ and SOTA. We consider the established leave-one-out protocol on \FAVC. Performance in \%. Abbreviations: rV – real Video, rA – real Audio, fA – fake Audio, W2L – Wav2Lip.}
\label{tab:FAVC_established_loo}
\centering
\resizebox{\linewidth}{!}{%
\begin{tabular}{@{}lccrrrrrrrrr@{}}
\toprule
\multicolumn{1}{c}{\multirow{2}{*}{Model}} & \multirow{2}{*}{Modality} & \multicolumn{2}{c}{rV+fA}                           & \multicolumn{2}{c}{W2L+rA}                       & \multicolumn{2}{c}{faceswap+fA}                  & \multicolumn{2}{c}{fsgan+fA}                     & \multicolumn{2}{c}{W2L+fA}                       \\
\multicolumn{1}{c}{}                       &                           & AP                        & \multicolumn{1}{c}{AUC} & \multicolumn{1}{c}{AP} & \multicolumn{1}{c}{AUC} & \multicolumn{1}{c}{AP} & \multicolumn{1}{c}{AUC} & \multicolumn{1}{c}{AP} & \multicolumn{1}{c}{AUC} & \multicolumn{1}{c}{AP} & \multicolumn{1}{c}{AUC} \\ \midrule
LipForensics~\cite{LipForensics}           & V                         & -                         & \multicolumn{1}{c}{-}   & 100.00                 & 100.00                  & 99.98                  & 99.94                   & 100.00                 & 99.99                   & 99.95                  & 99.84                   \\
RealForensics~\cite{RealForensics}         & V                         & -                         & \multicolumn{1}{c}{-}   & 99.97                  & 99.87                   & 99.98                  & 99.91                   & 99.99                  & 99.94                   & 99.70                  & 99.21                   \\ \midrule
AVAD~\cite{avad}                           & AV                        & \multicolumn{1}{r}{62.40} & 71.60                   & 93.60                  & 93.70                   & 95.30                  & 95.80                   & 94.10                  & 94.30                   & 93.80                  & 94.10                   \\
AVFF~\cite{AVFF}                           & AV                        & \multicolumn{1}{r}{93.30} & 92.40                   & 94.80                  & 98.20                   & 100.00                 & 100.00                  & 99.90                  & 100.00                  & 99.40                  & 99.80                   \\ \midrule
SIMBA binary                               & AV                        & \multicolumn{1}{r}{99.79} & 98.76                   & 91.20                  & 95.63                   & 100.00                 & 100.00                  & 100.00                 & 100.00                  & 100.00                 & 100.00                  \\
SIMBA multiclass                           & AV                        & \multicolumn{1}{r}{99.84} & 98.91                   & 94.71                  & 96.88                   & 100.00                 & 100.00                  & 100.00                 & 100.00                  & 100.00                 & 100.00                  \\ \bottomrule
\end{tabular} %
} 
\end{table*}

\section{Family Split Performance}
\label{supp:family_split_results}

\begin{figure*}[t]
    \begin{subfigure}{.5\textwidth}
        \centering
        \includegraphics[width=\textwidth]{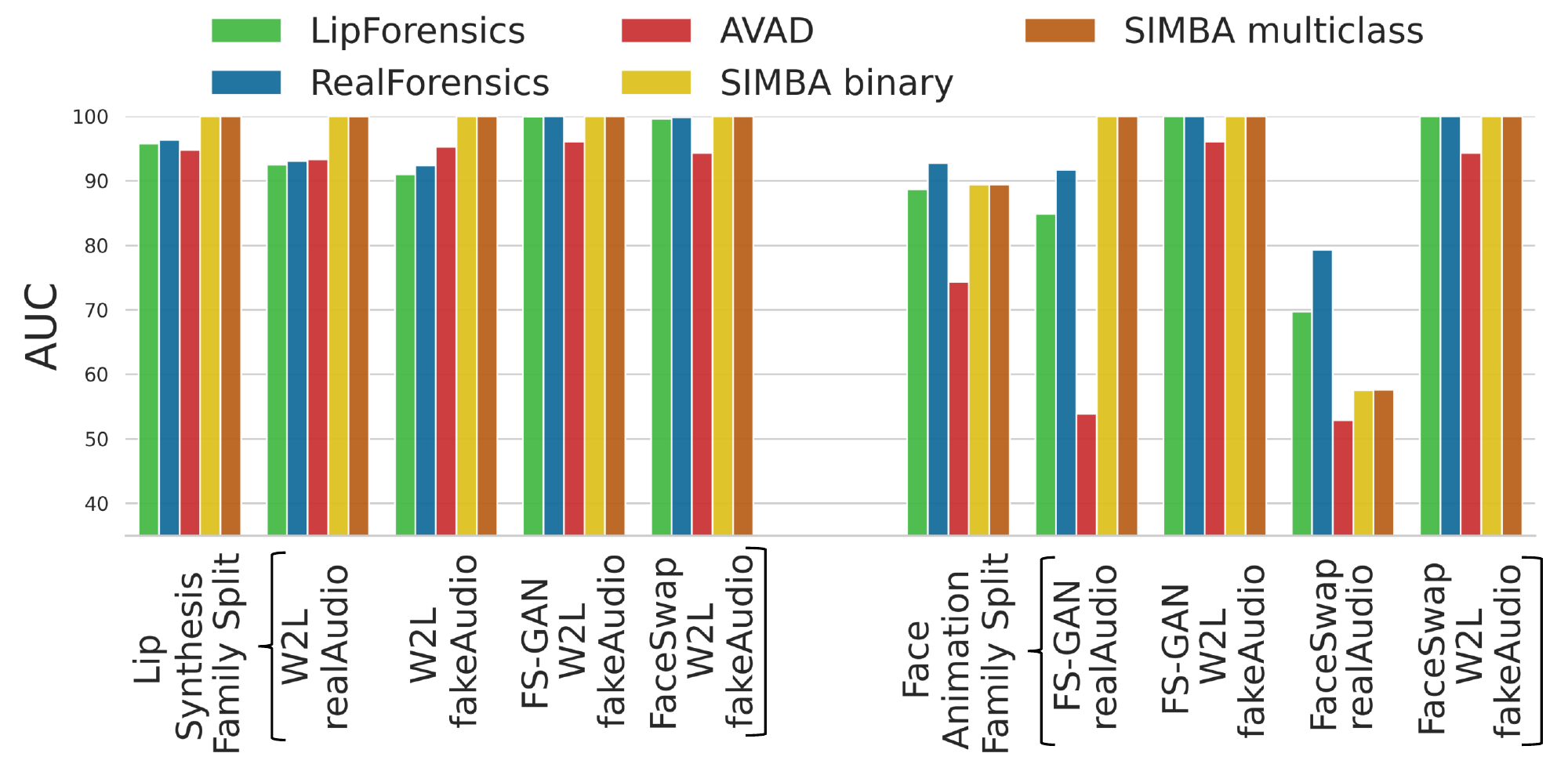}
        \caption{\FAVC}
        \label{fig:family_split_favc}
    \end{subfigure}%
    \begin{subfigure}{.5\textwidth}
        \centering
        \includegraphics[width=\textwidth]{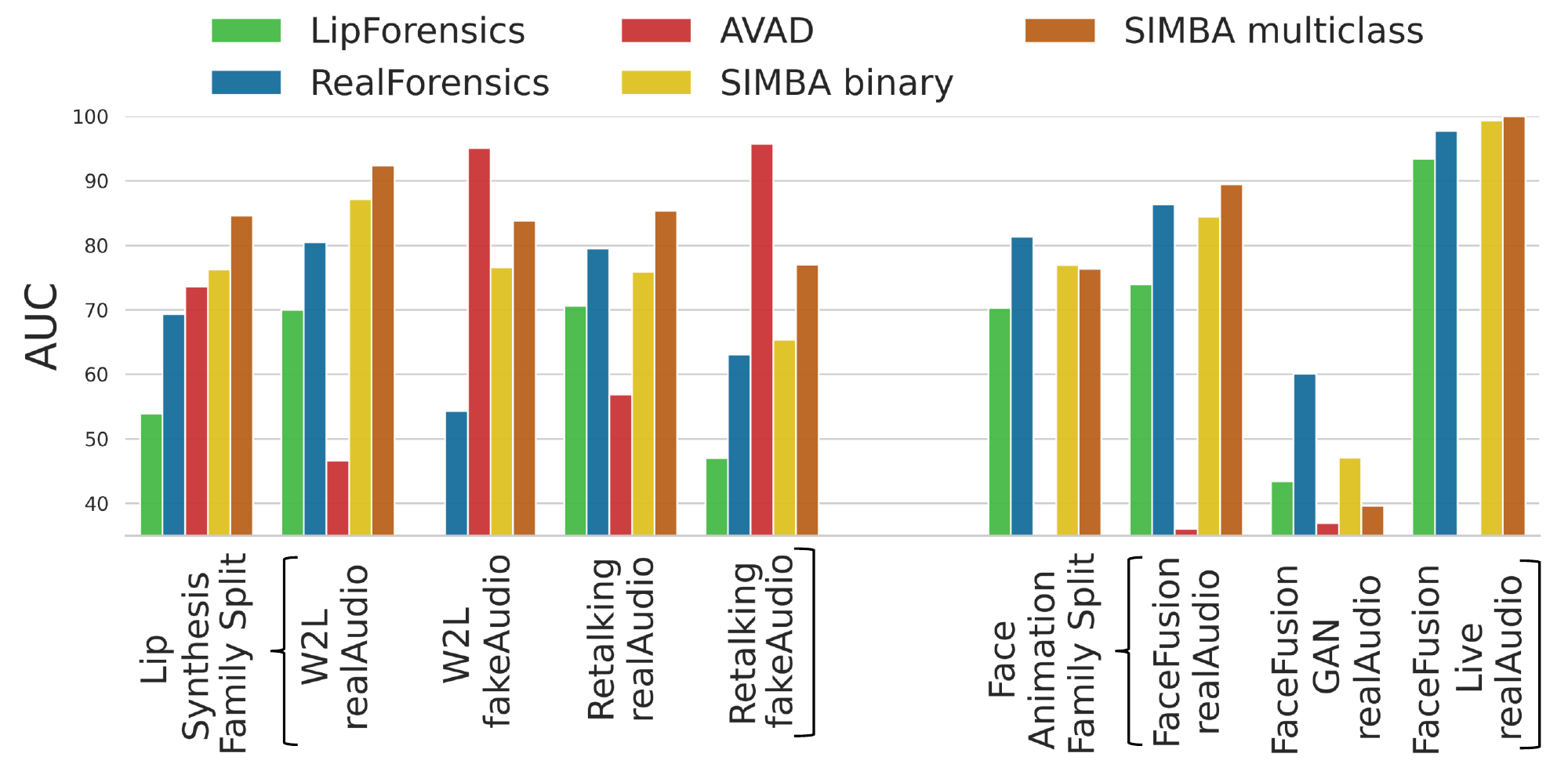}
        \caption{\DS}
        \label{fig:family_split_ds}
    \end{subfigure}
    \caption{Cross-manipulation comparison using the proposed family splits (as AUC).}
    \label{fig:family_split}
\end{figure*}

Figure~\ref{fig:family_split} shows the performance of SOTA models~\cite{LipForensics, RealForensics, avad} and \ourname\ models trained with subsampling and temporal jittering on our family splits introduced in Section~\ref{sec:evaluation_protocols}. Generalizing to the Lip Synthesis Family split results in overall high performance on \FAVC. Especially \ourname\ generalizes perfectly. AVAD achieves slightly higher performance when the audio is fake, suggesting that it slightly over-relies on the auditory modality. The unimodal models generalize perfectly when they face a face animation manipulation together with \wl\ but reach slightly lower results on only \wl. \emph{The combination of multiple manipulations seems to amplify the ``fake'' signal from which the models benefit.}
This finding is also confirmed on the Face Animation Family Split, which shows perfect scores for the supervised models on \FSGAN/\FS\ + \wl. AVAD struggles with real audio, which was already found in the method split (Sec.~\ref{sec:cross_manipulation_results}). We can also confirm from the method split that \FS\ is the hardest split for all models. Even though it has artifacts easily detectable by humans, it seems these artifacts are not shared with any other manipulation, making it much harder to generalize to. Overall, the performance is lower on the Face Animation Family than the Lip Synthesis Family Split, revealing that \emph{it is more challenging to generalize from lip synthesis manipulation to face animation manipulations than vice versa}.

Family Split results on \DS\ are visualized in Figure~\ref{fig:family_split_ds}. Scores drop significantly on the Lip Synthesis Family Split on \DS\ compared to \FAVC. This suggests that \FS\ and \FSGAN\ have more in common with \wl\ than the more recent FaceFusion manipulations. AVAD's alignment focus helps detect fake audio splits, as it surpasses all other models on these. 
\ourname\ models outperform the unimodal models on the Lip Synthesis Split even though it was trained only on real audio, thus, has no benefit out of its multimodality. Already the visual branch of \ourname\ provides generalization capabilities. 
On the face animation family split, \ourname\ is outperformed by RealForensics. This split is purely real audio, and it seems the fake audio during training hurts \ourname s generalization capabilities. FaceFusion GAN is the hardest split to generalize to, whereas FaceFusion Live is the easiest, which is the same finding as for the method split (Sec.~\ref{sec:cross_manipulation_results}). On average, performance is much lower on \DS\ and on \FAVC, showing that \emph{the more recent manipulations of \DS\ present a harder generalization challenge for SOTA models.}

Finally, the family split performance is lower than the method split results, highlighting that \emph{the more realistic family splits actually pose a greater generalization challenge for SOTA models.}

\section{Ablation of the Sampling Strategy During Evaluation}
\label{supp:ablation_eval_sampling}
As different SOTA models use different sampling strategies when evaluating, we investigate three different sampling strategies for \ourname\ in Table~\ref{tab:ablation_eval_sampling_strategy}. First, we sample one clip starting from the beginning. Second, multiple clips are sampled from the video, and the prediction scores are aggregated via the mean operation (\textit{clips mean}). Last, multiple clips are sampled and aggregated with the max operation (\textit{clips max}). Each clip is 80 frames long (subsampling stepsize * number of frames: $5*16=80$). When multiple clips are sampled per video, as many clips are sampled with no overlap as can be extracted from the entire video. Yet, the maximum number of clips per video is set to five. We found out that aggregating multiple clips hurts the generalization performance. Consequently, we report results with the \emph{beginning} schema in the paper.

\begin{table}[ht]
\caption{Evaluation sampling strategy analysis of \DS\ via a cross-manipulation leave-one-out comparison of \ourname\ binary and multiclass trained with subsampling and temporal jittering. Performance is given as AUC.}
\label{tab:ablation_eval_sampling_strategy}
\resizebox{\textwidth}{!}{%
\begin{tabular}{@{}lrrrrrr@{}}
\toprule
                                       & \multicolumn{1}{l}{AVG} & \multicolumn{1}{l}{\begin{tabular}[c]{@{}l@{}}Wav2Lip\\ Split\end{tabular}} & \multicolumn{1}{l}{\begin{tabular}[c]{@{}l@{}}Retalking\\ Split\end{tabular}} & \multicolumn{1}{l}{\begin{tabular}[c]{@{}l@{}}FaceFusion\\ Split\end{tabular}} & \multicolumn{1}{l}{\begin{tabular}[c]{@{}l@{}}FaceFusion\\ GAN Split\end{tabular}} & \multicolumn{1}{l}{\begin{tabular}[c]{@{}l@{}}FaceFusion\\ Live Split\end{tabular}} \\ \midrule
Binary beginning              & \textbf{93.59}                   & 99.10                                                                       & 93.80                                                                         & 96.51                                                                          & 78.64                                                                              & 99.89                                                                               \\
Binary clips mean     & 91.02                   & 99.15                                                                       & 92.27                                                                         & 94.30                                                                          & 69.54                                                                              & 99.83                                                                               \\
Binary clips max      & 90.47                   & 98.81                                                                       & 90.60                                                                         & 93.85                                                                          & 69.23                                                                              & 99.86                                                                               \\ \midrule
Multiclass beginning            & \textbf{93.06}                   & 99.51                                                                       & 92.44                                                                         & 95.26                                                                          & 78.41                                                                              & 99.70                                                                               \\
Multiclass clips mean & 90.78                   & 98.98                                                                       & 89.10                                                                         & 92.61                                                                          & 73.60                                                                              & 99.63                                                                               \\
Multiclass clips max  & 91.28                   & 99.03                                                                       & 90.15                                                                         & 92.99                                                                          & 74.56                                                                              & 99.66                                                                              \\ \bottomrule
\end{tabular}%
}
\end{table}

\section{Visualizing SIMBA's Embedding Spaces}
\label{supp:embeddings}
\begin{figure}[ht]
	\centering
    \begin{subfigure}[b]{0.49\linewidth}
        \centering
        \includegraphics[width=\textwidth]{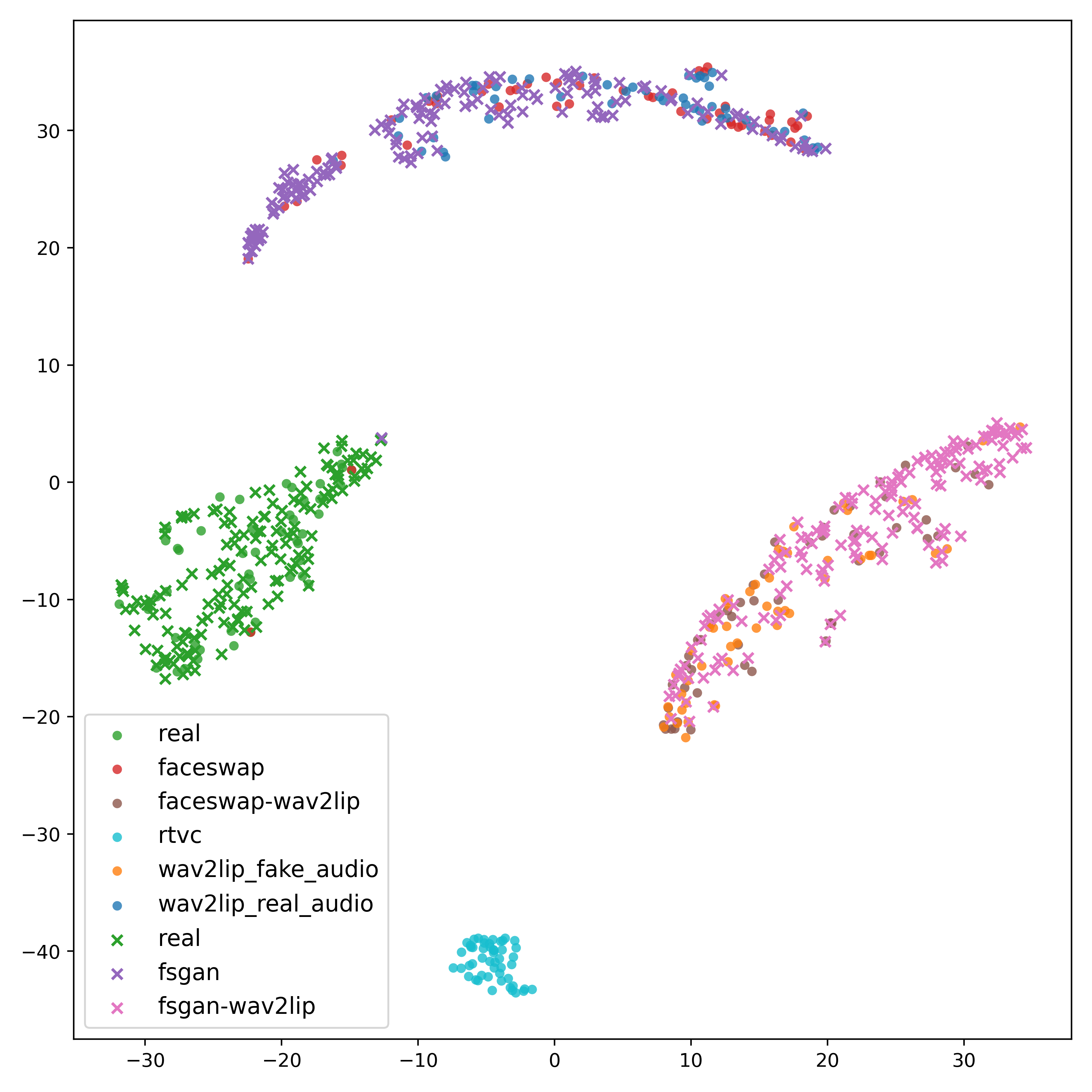}
        \caption{\ourname\ binary \FSGAN}
        \label{fig:tsne_favc_binary_fsgan}
    \end{subfigure}
    \begin{subfigure}[b]{0.49\linewidth}
        \centering
        \includegraphics[width=\textwidth]{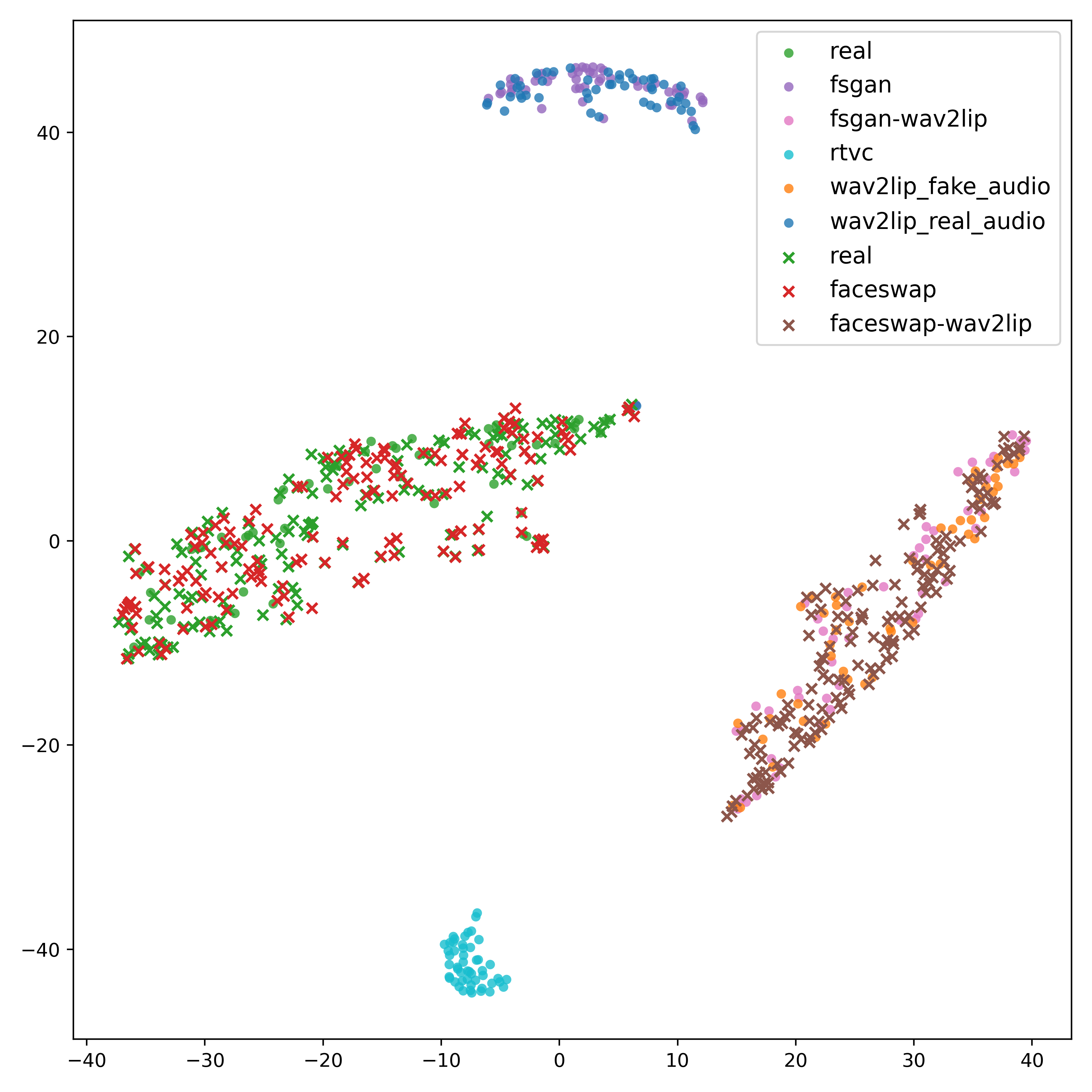}
        \caption{\ourname\ binary \FS}
        \label{fig:tsne_favc_binary_faceswap}
    \end{subfigure}
    \begin{subfigure}[b]{0.49\linewidth}
        \centering
        \includegraphics[width=\textwidth]{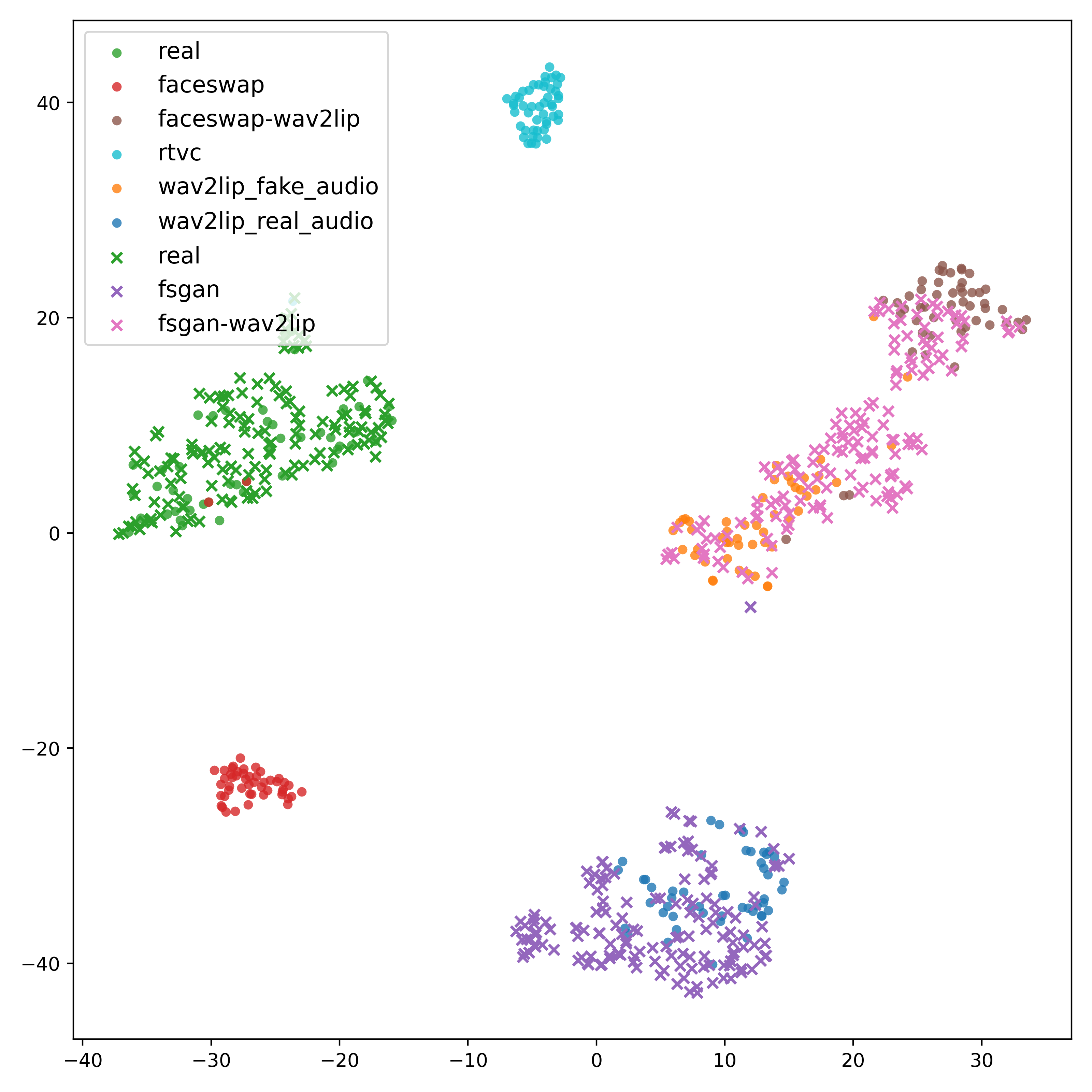}
        \caption{\ourname\ multiclass \FSGAN}
        \label{fig:tsne_favc_multiclass_fsgan}
    \end{subfigure}
    \begin{subfigure}[b]{0.49\linewidth}
        \centering
        \includegraphics[width=\textwidth]{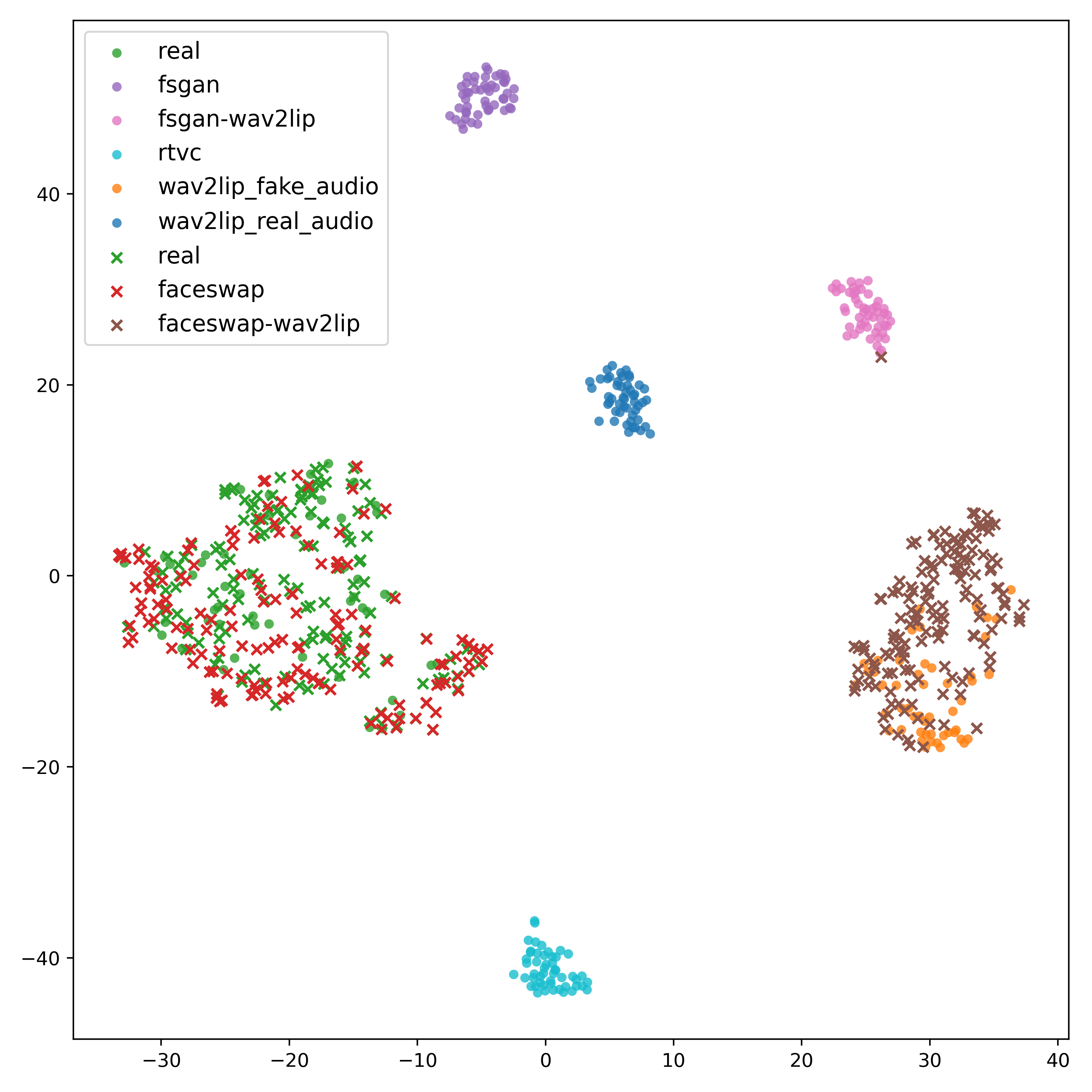}
        \caption{\ourname\ multiclass \FS}
        \label{fig:tsne_favc_multiclass_faceswap}
    \end{subfigure}
	\caption{Visualization of the embedding space for \ourname\ binary and multiclass on \FAVC. The left column shows \ourname\ models trained on the \FSGAN\ method split, whereas the \FS\ method split was used for the right column. $\circ$ show manipulations seen during training and $\times$ the unseen manipulations.}
	\label{fig:tsne_favc}
\end{figure}

Figure~\ref{fig:tsne_favc} visualizes the embedding space of \ourname\ binary and \ourname\ multiclass on the \FSGAN\ and \FS\ method split of \FAVC\ using t-distributed stochastic neighbor embedding (t-SNE)\footnote{Geoffrey E. Hinton, and Sam Roweis. Stochastic neighbor embedding. In \textit{Advances in neural information processing systems}, 15, 2002.}.
The figure plots the in-distribution manipulations (seen during training) as $\circ$ and the out-of-distribution samples (unseen manipulations) as $\times$.

Notice that the multiclass models form clear, distinct clusters for each in-distribution manipulation type (Fig.~\ref{fig:tsne_favc_multiclass_fsgan}, \ref{fig:tsne_favc_multiclass_faceswap}). In contrast to that, \wl\ fake audio overlaps with \FS+\wl\ and \wl\ real audio overlaps with \FS\ for \ourname\ binary on the \FSGAN\ split (Fig.~\ref{fig:tsne_favc_binary_fsgan}). Similarly, \wl\ fake audio overlaps with \FSGAN+\wl\ and \wl\ real audio overlaps with \FSGAN\ for \ourname\ binary on the \FS\ split (Fig.~\ref{fig:tsne_favc_binary_faceswap}). Notice that these overlaps are audio-related. \emph{The binary models form real-video-real-audio, real-video-fake-audio, fake-video-real-audio, and fake-video-fake-audio clusters.}

Unseen manipulations during training are aligned to existing clusters during inference. The \ourname\ binary model on the \FSGAN\ split maps nicely the \FSGAN\ samples to the \wl\ real audio / \FS\ cluster, the \FSGAN+\wl\ samples to the \wl\ fake audio / \FS+\wl\ cluster, and the real samples to the real cluster. The corresponding multiclass model (Fig.~\ref{fig:tsne_favc_multiclass_fsgan} maps the \FS+\wl\ samples to \wl\ fake audio and to the \FS+\wl cluster and the real samples to the real cluster. Yet, the \FSGAN\ samples are aligned with the \wl\ real audio cluster and not with the \FS\ cluster. This reveals that \emph{for \ourname, \FSGAN\ has more learned artifacts in common with the lip synthesis manipulation \wl\ than with the face animation manipulation \FS}. Still, no manipulation overlaps with the real cluster, resulting in the almost perfect generalization performance of \ourname\ in Figure~\ref{fig:method_split_favc}.

The \FS\ real audio manipulation is the hardest manipulation to generalize to for \ourname\ (Fig.~\ref{fig:method_split_favc}). The embedding spaces of \ourname\ on the \FS\ split show a high overlap between the \FS\ real audio and real cluster. The \FS+\wl\ fake audio is perfectly aligned with the fake-video-fake-audio cluster (Fig.~\ref{fig:tsne_favc_binary_faceswap}). The multiclass \ourname\ aligns the unseen \FS+\wl\ with the \wl\ fake audio and not with the \FSGAN+\wl\ cluster. This shows that \emph{the \wl\ part of the combined manipulation has a greater impact on the final decision than the face animation part for \ourname}. \ourname s performance drops when generalizing to the \FS\ real audio manipulation compared to the \FSGAN\ real audio manipulation can be explained by the multiclass embedding spaces. \emph{Since \FSGAN\ and \FS\ do not share artifacts (for \ourname), and only \FSGAN\ shares artifacts with \wl, the unseen \FS\ real audio manipulation is not aligned with any seen manipulation but with the real cluster.}

\begin{figure}[ht]
	\centering
    \begin{subfigure}[b]{0.49\linewidth}
        \centering
        \includegraphics[width=\textwidth]{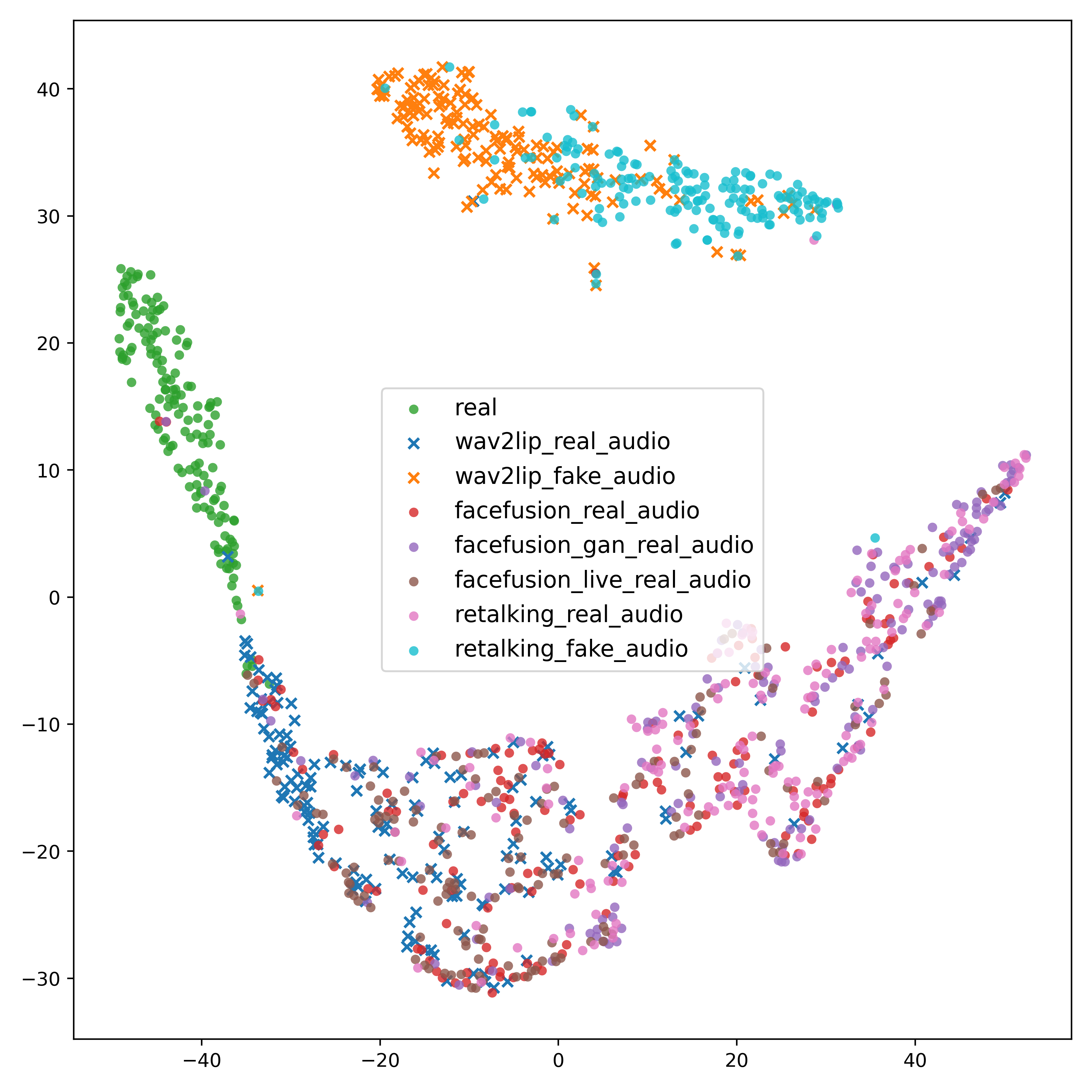}
        \caption{\ourname\ binary \wl}
        \label{fig:tsne_ds_binary_wav2lip}
    \end{subfigure}
    \begin{subfigure}[b]{0.49\linewidth}
        \centering
        \includegraphics[width=\textwidth]{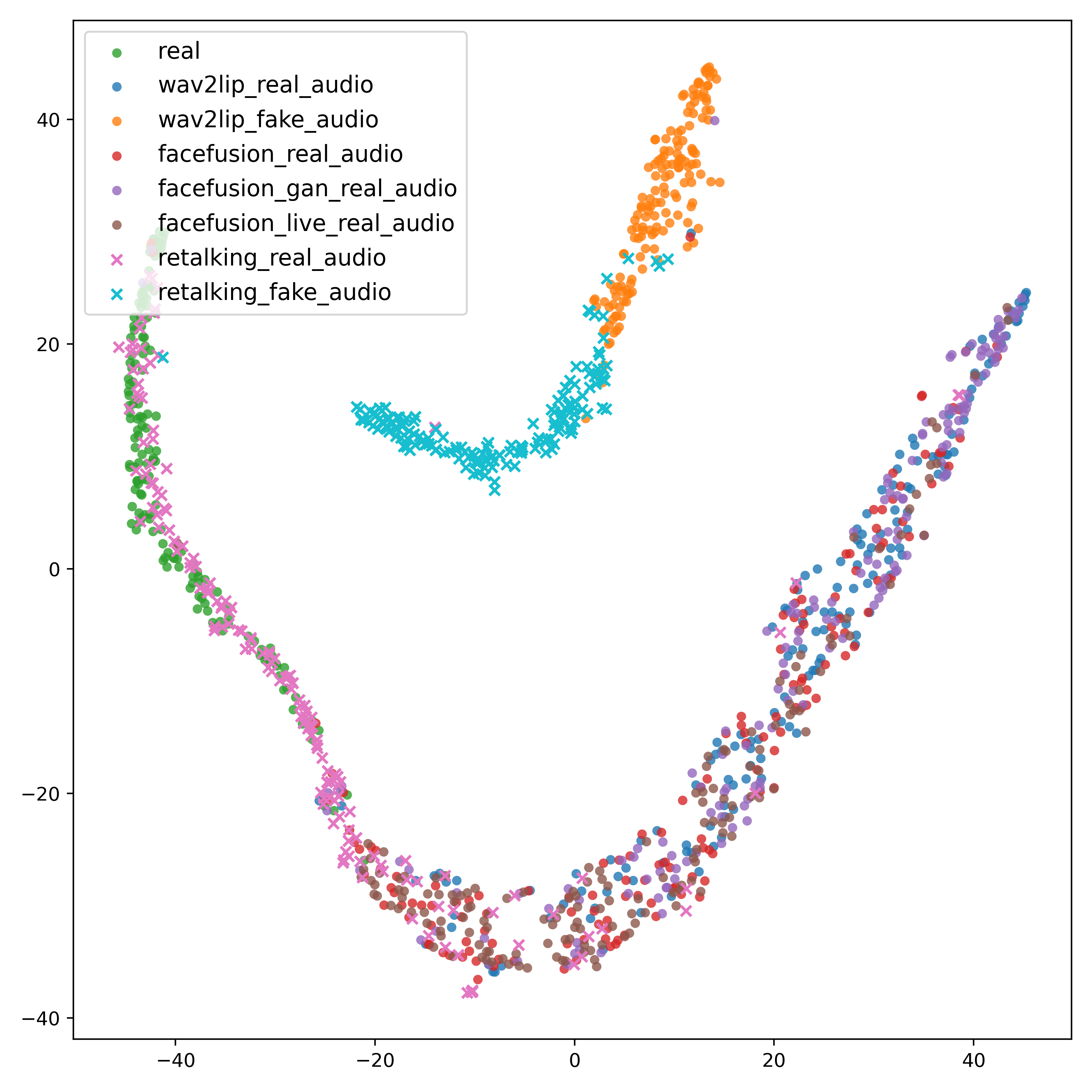}
        \caption{\ourname\ binary Retalking}
        \label{fig:tsne_ds_binary_retalking}
    \end{subfigure}
    \begin{subfigure}[b]{0.49\linewidth}
        \centering
        \includegraphics[width=\textwidth]{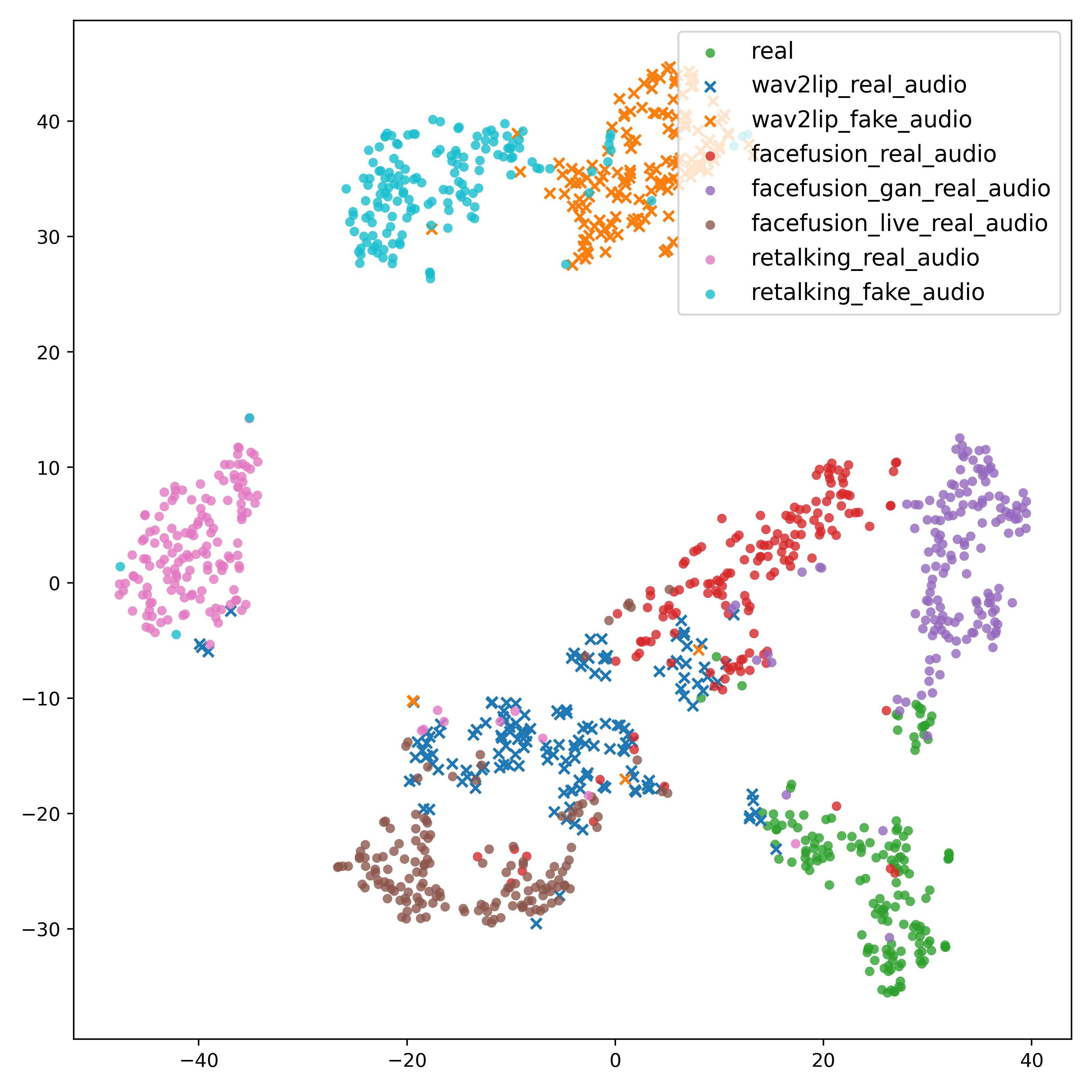}
        \caption{\ourname\ multiclass \wl}
        \label{fig:tsne_ds_multiclass_wav2lip}
    \end{subfigure}
    \begin{subfigure}[b]{0.49\linewidth}
        \centering
        \includegraphics[width=\textwidth]{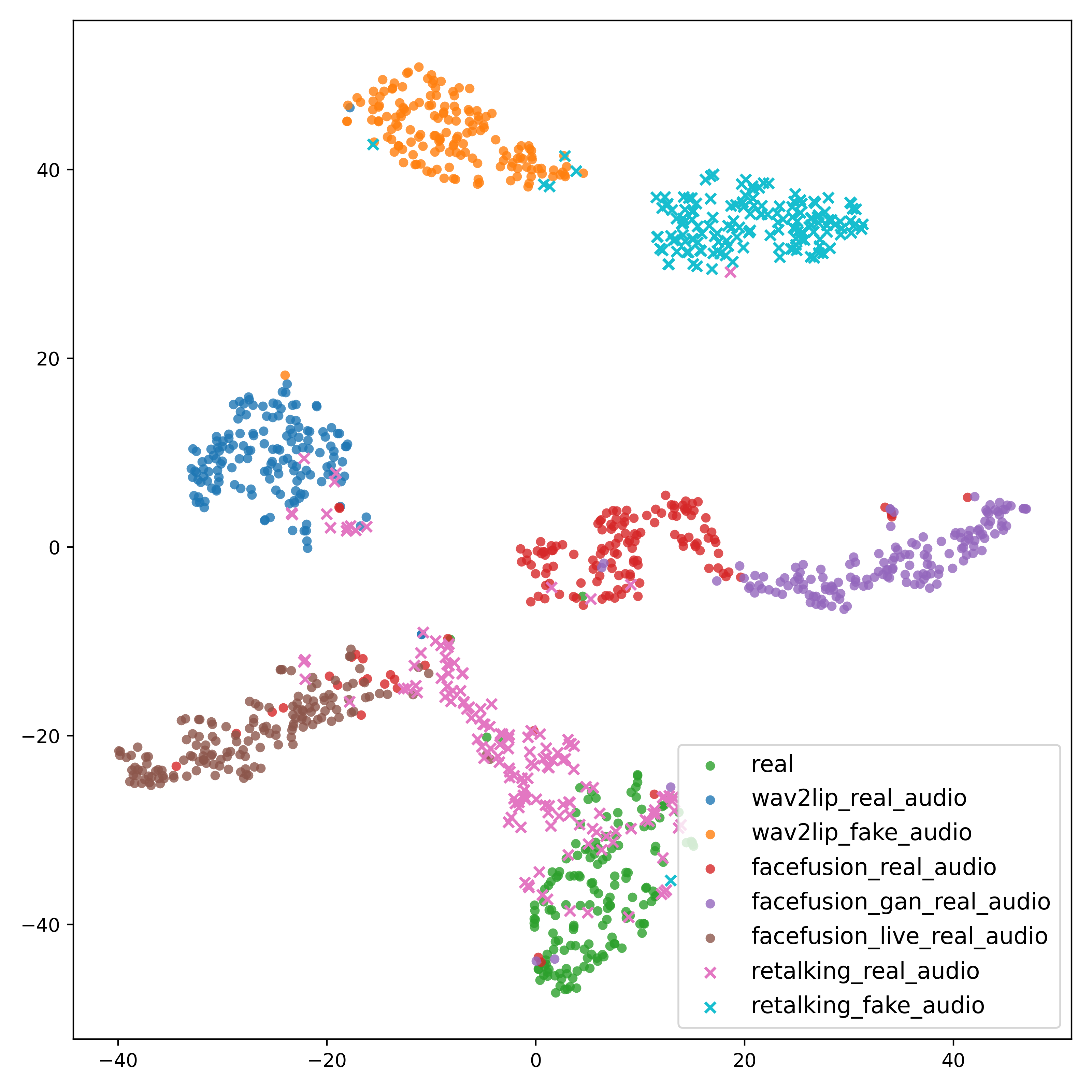}
        \caption{\ourname\ multiclass Retalking}
        \label{fig:tsne_ds_multiclass_retalking}
    \end{subfigure}
	\caption{Visualization of the embedding space for \ourname\ binary and multiclass on \DS. The left column shows \ourname\ models trained on the \wl\ method split, whereas the Retalking method split was used for the right column. $\circ$ show manipulations seen during training and $\times$ the unseen manipulations.}
	\label{fig:tsne_ds}
\end{figure}

Figure~\ref{fig:tsne_ds} displays the embedding spaces of \ourname\ binary and \ourname\ multiclass on the \wl\ and Retalking method split of \DS. As \DS\ does not have a real-video-fake-audio combination, the binary models form roughly three clusters (real-video-real-audio, fake-video-real-audio, and fake-video-fake-audio). \ourname\ binary on the \wl\ split aligns \wl\ real audio with the fake-video-real-audio cluster and \wl\ fake audio with the fake-audio cluster, resulting in almost perfect generalization performance as discussed in Section~\ref{sec:cross_manipulation_results}. The same holds for \ourname\ multiclass on the \wl\ split (Fig.~\ref{fig:tsne_ds_multiclass_wav2lip}). Yet, \wl\ fake audio forms a new cluster close to the Retalking clusters, and \wl\ real audio samples are aligned with the FaceFusion Live cluster, suggesting that \emph{the \wl\ real audio lip synthesis manipulation has more in common with the face animation manipulation FaceFusion Live than with the Retalking lip manipulation}. 

When generalizing to the Retalking split, \ourname s performance is slightly lower than on the \wl\ split (Fig.~\ref{fig:method_split_ds}). This is caused by the slight overlap between the Retalking real audio samples and the real cluster for \ourname\ binary as well as multiclass. 
Comparing the embedding spaces from \FAVC\ to the ones on \DS\ reveals that some unseen \DS-manipulations form clusters which do not match existing clusters. Unseen \FAVC-samples always show a high overlap with existing clusters. This suggests that \emph{the more recent manipulation techniques generate more dissimilar artifacts than the older manipulation techniques present in \FAVC}.

\end{document}